\documentclass[usenatbib,usegraphicx,useAMS]{mn2e}
\usepackage{amsmath}

\def\cf{{cf.}}
\def\eg{{e.g.}}
\def\ie{{i.e.}}

\title{Covariant Magnetoionic Theory I: Ray Propagation}
\author{Avery Broderick \& Roger Blandford}

\begin{document}
\maketitle

\begin{abstract}
Accretion onto compact objects plays a central role in high energy
astrophysics. In these environments, both general relativistic and
plasma effects may have significant impacts upon the propagation
of photons. We present a full general relativistic magnetoionic
theory, capable of tracing rays in the geometric optics approximation
through a magnetised plasma in the vicinity of a compact object. We
consider both the cold and warm, ion and pair plasmas.  When plasma
effects become large the two plasma eignemodes follow different ray
trajectories resulting in a large observable polarisation.  This has
implications for accreting systems ranging from pulsars and X-ray
binaries to AGN.
\end{abstract}


\section{Introduction} \label{Intro}
A considerable amount of effort has been invested in attempting to reproduce
the spectral properties of accreting compact objects.  A great
deal of this work has been concerned with fitting the unpolarised flux with
an underlying, physically motivated model of an accretion flow 
\citep[see \eg][]{Blan-Begl:99,Nary-Yi:94,Quat-Gruz:00}. These models have met
with some success, even being able to make testable predictions regarding
the accretion environment \citep[see \eg][]{Nary-Maha-Grin-Poph-Gamm:98}.
Because these models are primarily concerned with the physical structure of
the accretion flow, they ignore the effects that the combination of
dispersion and general relativity will have upon the spectra. Far from the
compact object this may not matter ($r \gg M$). However, for the emission
originating from near the compact object, this combination can be crucial.

General relativistic vacuum propagation effects have been extensively studied
in both the polarised and unpolarised cases. In some systems gravitational
lensing has been shown to have detectable consequences in certain regions
of the spectrum.  For example, \cite{Falc-Meli-Agol:00} have argued that the
black hole in the Galactic centre may be imaged directly at millimeter
wavelengths as a result of gravitational lensing. In addition, general
relativity has been  shown to have a depolarising influence upon photons
passing near the compact object
\citep[see \eg][]{Laor-Netz-Pira:90,Agol:97,Conn-Star-Pira:80}.
However, the fact that these studies ignore plasma effects make them
inapplicable to thick disks and at frequencies near the plasma and/or
cyclotron frequencies.

On the other hand, astrophysical plasma effects have also been studied in
detail, although
primarily in the context of nondispersive propagation effects upon the
polarisation, \eg~Faraday rotation and conversion
\citep[see \eg][]{Sazo-Tsyt:68,Sazo:69,Jone-ODel:77b,Jone-ODel:77a,Rusz-Bege:02}.
Weak dispersion has been considered in the form of scintillation
\citep[see \eg][]{Macq-Melr:00}. While this can lead to a high degree of
polarisation variability, it has a vanishing time/spatially averaged
value and doesn't otherwise affect spectral properties. In contrast,
strongly dispersive plasma effects have been extensively studied in the context
of radio waves propagating through the ionosphere \citep[see \eg][]{Budd:64}.
Here it has been found that dispersive plasma effects can play an important
role in determining the intensity and limiting polarisation of the radio
waves. None the less, neither of these type of plasma effects have been
studied in conjunction with general relativistic effects.

There have been some attempts at treating both general relativistic and plasma
effects. However, these have been restricted to either unmagnetised plasmas
\citep[see \eg][]{Kuls-Loeb:92}, or to nondispersive emission effects
\citep[see \eg][]{Brom-Meli-Liu:01}. Both of these have only limited
applicability for realistic accretion flows.

We present a fully general relativistic magnetoionic theory. This is a
natural extension of the previous work combining both general relativistic and
plasma effects upon wave propagation in the geometrical optics limit. This
will be presented in four sections with \S\ref{Theory} developing the theory,
 \S\ref{Examples} presenting some simple example applications, and
\S\ref{Conclusion} containing conclusions.  Throughout this paper the
($-+++$) metric signature will be used, and $\hbar=c=1$.

In a subsequent paper II we will discuss the details of performing radiative
transfer in general relativistic plasma environments.  These have been expressly
neglected here in the interest of clarity.

\section{Theory} \label{Theory}
The natural place to begin a study of plasma modes is the covariant
formulation of Maxwell's equations \citep[see \eg][]{Misn-Thor-Whee:73}:
\begin{equation}
\nabla_\mu F^{\nu \mu} = 4 \pi J^\nu
\;\; \mbox{and} \;\;
\nabla_\mu \mbox{}^*\!F^{\nu \mu} = 0 \,,
\label{maxwell1}
\end{equation}
where $F^{\nu \mu} \equiv \nabla^\nu A^\mu - \nabla^\mu A^\nu$ is the
electromagnetic field tensor,
$^*\!F^{\nu \mu} \equiv \frac12 \varepsilon^{\nu \mu \alpha \beta}
F_{\alpha \beta}$ is the dual to $F^{\mu \nu}$
($\varepsilon^{\mu\nu\alpha\beta}$ is the Levi-Civita pseudo tensor)
, and $J^\nu$ is the current
four-vector.  In order to close this set of equations, a relation between
the current and the electromagnetic fields is required.  For the field
strengths of interest here, this will take the form of Ohm's Law:
\begin{equation}
J^\nu = \sigma^\nu_{~\mu} F^{\mu \alpha} \overline{u}_\alpha \,,
\label{ohms_law}
\end{equation}
where $\overline{u}^\mu$ is the average plasma four-velocity and
$\sigma^\nu_{~\mu}$ is the covariant generalisation of the
conductivity tensor, defined by this relationship. As a result of
the anti-symmetry of $F^{\mu \nu}$, the conductivity will in general
have only nine physically meaningful components, namely the spatial
components in the slicing orthogonal to $\overline{u}^\mu$.
Nonetheless, in order to investigate the behaviours of plasma modes in
a general relativistic environment, it is necessary to express the
conductivity in this covariant fashion.

This can be more naturally expressed in terms of
$E^\mu \equiv F^{\mu \nu} \overline{u}_\nu$ and $B^\mu \equiv~^*\!F^{\mu \nu} \overline{u}_\nu$,
the four-vectors coincident with the electric and magnetic field vectors
in the locally flat centre-of-mass rest (LFCR) frame of the plasma.  In terms
of $E^\mu$ and $B^\mu$, the electromagnetic field tensor and its dual take the
forms
\begin{align}
F^{\mu \nu} &= \overline{u}^\mu E^\nu - E^\mu \overline{u}^\nu + 
\varepsilon^{\mu\nu\alpha\beta}\,\overline{u}_\alpha\,B_\beta \,,
\label{fmunua} \\
^*\!F^{\mu \nu} &= \overline{u}^\mu B^\nu - B^\mu \overline{u}^\nu + 
\varepsilon^{\mu\nu\alpha\beta}\,\overline{u}_\alpha\,E_\beta \,.
\label{fmunub}
\end{align}
Inserting these and Ohm's law into Maxwell's equations yields eight
partial differential equations,
\begin{align}
\nabla_\mu  \left( \overline{u}^\nu E^\mu - E^\nu \overline{u}^\mu + 
\varepsilon^{\nu\mu\alpha\beta}\,\overline{u}_\alpha\,B_\beta \right) &=
4 \pi \sigma^\nu_{~\mu} E^\mu \,, \label{maxwell2a} \\
\nabla_\mu  \left( \overline{u}^\nu B^\mu - B^\nu \overline{u}^\mu + 
\varepsilon^{\nu\mu\alpha\beta}\,\overline{u}_\alpha\,E_\beta \right) &=
0 \label{maxwell2b} \,,
\end{align}
which may be solved for $E^\mu$ and $B^\mu$ given an explicit form of the
conductivity.

\subsection{Geometric Optics Approximation} \label{GOA}
The general case can be prohibitively difficult to solve for physically
interesting plasmas.  Fortunately, the problem can be significantly simplified
by making use of a two length scale expansion (also known as the WKB, Eikonal,
or Geometric Optics approximations) in terms of $\lambda/{\cal L}$,
where $\lambda$ and ${\cal L}$ are the wavelength and typical plasma
length scale, respectively.  In this approximation it is assumed that the
electric and magnetic fields have a slowly varying amplitude with a rapidly
varying phase, \ie~$E^\mu,\,B^\mu \propto \exp\left(iS\right)$ where
$S$ is the action, and $\nabla_\mu S = k_\mu$ defines the wave four-vector.
Then, to first order in $\lambda/{\cal L}$, Maxwell's equations are
\begin{align}
k_\mu  \left( \overline{u}^\nu E^\mu - E^\nu \overline{u}^\mu + 
\varepsilon^{\nu\mu\alpha\beta}\,\overline{u}_\alpha\,B_\beta \right) &=
4 \pi \sigma^\nu_{~\mu} E^\mu \,, \label{maxwell3a} \\
k_\mu  \left( \overline{u}^\nu B^\mu - B^\nu \overline{u}^\mu + 
\varepsilon^{\nu\mu\alpha\beta}\,\overline{u}_\alpha\,E_\beta \right) &=
0 \label{maxwell3b} \,.
\end{align}

At this point it is useful to point out a number of properties of $E^\mu$ and
$B^\mu$ that follow directly from their definitions and Maxwell's equations.
\begin{description}
\item[({\em i})] $\overline{u}_\mu E^\mu = \overline{u}_\mu B^\mu = 0$, which follows directly from
the definitions of $E^\mu$ and $B^\mu$ and the antisymmetry of $F^{\mu\nu}$ and
$^*\!F^{\mu\nu}$.
\item[({\em ii})] $k_\mu B^\mu = 0$, which follows from equation
(\ref{maxwell3b}) and the definition of $B^\mu$.
\item[({\em iii})] $E_\mu B^\mu = 0$, which follows from
$\omega E_\mu B^\mu = E_\mu k_\nu\mbox{}^*\!F^{\mu\nu} = 0$, where
$\omega \equiv -k_\mu \overline{u}^\mu$ (chosen so that $\omega$ is positive)
is the frequency in the LFCR frame and is assumed to be nonzero.
\item[({\em iv})] $\omega B^\mu B_\mu =
- \varepsilon^{\mu\nu\alpha\beta} B_\mu k_\nu \overline{u}_\alpha E_\beta$, which also
follows from equation (\ref{maxwell3b}), $\omega B^\mu B_\mu +
\varepsilon^{\mu\nu\alpha\beta} B_\mu k_\nu \overline{u}_\alpha E_\beta
= B_\mu k_\nu\mbox{}^*\!F^{\mu\nu} = 0$.
\end{description}
Properties ({\em i})-({\em iv}) define $B^\mu$ in terms of $k^\mu$, $E^\mu$,
and $\overline{u}^\mu$:
\begin{equation}
B^\mu = -\frac1\omega \varepsilon^{\mu\nu\alpha\beta} k_\nu \overline{u}_\alpha E_\beta\,.
\label{magnetic}
\end{equation}

Substituting equation (\ref{magnetic}) into equations (\ref{fmunua}) and
(\ref{fmunub}) gives
\begin{align}
F^{\mu \nu} &= \frac1\omega \left( k^\mu E^\nu - E^\mu k^\nu \right) \,,
\label{fmunu2a} \\
^*\!F^{\mu \nu} &= \frac1\omega 
\varepsilon^{\mu\nu\alpha\beta}\,k_\alpha\,E_\beta \,.
\label{fmunu2b}
\end{align}
Inserting these back into Maxwell's equations and combining yields
\begin{equation}
\Omega^\mu_{~\nu} E^\nu = 0 \,,
\label{disp_eq}
\end{equation}
where
\begin{equation}
\Omega^\mu_{~\nu} \equiv
\left( k^\alpha k_\alpha \delta^\mu_\nu - k^\mu k_\nu
- 4\pi i \omega \sigma^\mu_{~\nu} \right) \,
\label{disp_tensor} 
\end{equation}
defines the dispersion tensor.

Note that this is extremely general, all of the local physics is contained in
the conductivity tensor.  The expressions for the electromagnetic field
tensor and its dual are for the radiation fields only.  Hence, external
fields appear only in the conductivity.

\subsection{Ray Equations} \label{RE}
Rays are well defined in the context of geometric optics.  These are
curves which are orthogonal at every point to the surfaces of constant
phase ($S$).
Given a relation in the form of equation (\ref{disp_eq}) it is possible to
explicitly construct these rays.  This has been done in detail for Euclidean
spaces \citep[see \eg][]{Wein:62}.  The generalisation to a Riemannian
space is straightforward and will be done in analogy with \cite{Wein:62}.

Consider the general case of an equation governing the dynamics of a field,
$\Psi$, in space time in terms of a linear operator, $\mathbf M$,
\begin{equation}
{\mathbf M}\left( \nabla_\mu, x^\mu \right) {\Psi} = 0 \,.
\end{equation}
Expanding in a two length scale approximation, as in \S2.1, gives
to lowest order
\begin{equation}
{\mathbf M}\left( k_\mu, x^\mu \right) {\Psi} = 0 \,.
\end{equation}
This implies that
$\det {\mathbf M}\left(k_\mu,x^\mu\right)=0$ along the rays of the wave
field.  This provides a dispersion relation, $D\left(k_\mu,x^\mu\right)$, a
scalar function of the wave four-vector and position that vanishes along the
ray.  If the eigenvalues of ${\mathbf M}$ are nondegenerate, then this also
uniquely defines the polarisation of $\Psi$.

The ray can now be explicitly constructed by employing the least action
principle.  The action can be explicitly constructed from the wave four-vector
and the position by
\begin{equation}
S(\tau_1,\tau_2) = \int_{\tau_1}^{\tau_2}
k_\mu \frac{d x^\mu}{d \tau} d \tau \,,
\end{equation}
where $\tau$ is an affine parameter along the ray.
Let $\Gamma$ be the hypersurface passing through the point
$x^\mu\left(\tau_1\right)$.  By definition, $k_\mu\left(\tau_1\right)$
is perpendicular to $\Gamma$.  By varying $S\left(\tau_1,\tau_2\right)$ 
with respect to $k_\mu$ and $x^\mu$, restricting $x^\mu \left(\tau_1\right)$
to lie on $\Gamma$, it is possible to derive equations which define
the ray,
\begin{align}
\delta S &= \int_{\tau_1}^{\tau_2} \left[ \frac{dk_\mu}{dx^\nu}\,\delta x^\nu 
\frac{dx^\mu}{d\tau} + k_\mu \:\delta \!\!\left( \frac{dx^\mu}{d\tau} 
\right) \right] d\tau \\ 
&= \int_{\tau_1}^{\tau_2} \left[ \frac{dk_\mu}{dx^\nu} 
\frac{dx^\mu}{d\tau} - \frac{dk_\mu}{d\tau} \frac{dx^\mu}{dx^\nu} \right]
\delta x^\nu d\tau + k_\mu \:\delta x^\mu \Bigl|_{\tau_1}^{\tau_2} \,.
\nonumber
\end{align}
Because $x^\mu\left(\tau_1\right)$ is restricted to lie upon $\Gamma$, 
$k_\mu \delta x^\mu \bigl|_{\tau_1}=0$.  Because at $\tau_2$ it is necessary
for $k_\mu \left( \tau_2 \right) = \nabla_\mu S \rightarrow
\delta S = k_\mu\left(\tau_2\right)\delta x^\mu\left(\tau_2\right)$.
These imply that the integral must vanish for arbitrary variations.
This will be generally true if
\begin{equation}
\frac{dx^\mu}{d\tau} = \left( \frac{\partial D}{\partial k_\mu} \right)_{x^\mu}
\;\;\mbox{and}\;\;\;
\frac{dk_\mu}{d\tau} = -\left( \frac{\partial D}{\partial x^\mu} 
\right)_{k_\mu} \,,
\label{ray_eqs}
\end{equation}
and hence,
\begin{align}
\frac{dk_\mu}{dx^\nu} \frac{dx^\mu}{d\tau} 
&- \frac{dk_\mu}{d\tau} \frac{dx^\mu}{dx^\nu} = \nonumber \\
& \left( \frac{\partial D}{\partial k_\mu} \right)_{x^\mu} 
\frac{d k_\mu}{d x^\nu}
+\left( \frac{\partial D}{\partial x^\mu} \right)_{k_\mu} 
\frac{d x^\mu}{d x^\nu} \nonumber \\
&\quad = \frac{dD}{d\tau} \frac{d\tau}{d x^\nu} = 0\,, \nonumber
\end{align}
where the final equality follows from the fact that $D$ is constant along
the path (namely $D\left(k_\mu,x^\mu\right) = 0$). Therefore, equations
(\ref{ray_eqs}) can be used to construct a ray given initial conditions
and a dispersion relation. These are covariant analogues of Hamilton's
equations.  Note that the affine parameterisation depends upon the
particular form of the dispersion relation.  For example, from
$D'\left(k_\mu,x^\mu\right) \equiv f\left(k_\mu,x^\mu\right)
D\left(k_\mu,x^\mu\right)$ it is possible to construct the rays
associated with $D=0$, with the affine parameters related by
$d\tau' = d\tau/f$: \ie
\begin{align}
\frac{dx^\mu}{d\tau'}
&= \left( \frac{\partial D'}{\partial k_\mu} \right)_{x^\mu} \nonumber \\
&= f\left( \frac{\partial D}{\partial k_\mu} \right)_{x^\mu}
+ D\left( \frac{\partial f}{\partial k_\mu} \right)_{x^\mu} \nonumber \\
&= f \frac{dx^\mu}{d\tau} \,,
\end{align}
and similarly for $k_\mu$.  Hence, any convenient affine parameterisation can
be selected by employing the appropriate function $f$.

While this derivation is done in some generality, in this paper
${\mathbf M}=\Omega^\mu_{~\nu}$ and $\Psi=E^\mu$.

\subsection{Ohm's Law for Cold Plasmas} \label{cold}
At this point it is necessary to determine an explicit form for the 
conductivity tensor $\sigma^\mu_{~\nu}$.  For cold plasmas this can be
obtained via kinetic
theory.  Three assumptions are made in the derivations below; ({\em i}) the 
equations of motion of the electrons are well approximated by the lowest
order perturbations, ({\em ii}) the motions of the electrons
are non-relativistic, and ({\em iii}) the electrons execute motions over
a small enough region of space that all other forces may be considered
constant.  Assumptions ({\em i}) and ({\em ii}) are often employed in
standard plasma physics.
Assumption ({\em iii}) will generally be true as long as the geometric
optics approximation holds.

\subsubsection{Isotropic Cold Electron Plasma} \label{ICEP}
This is considered as an example and a limit of the case where a constant
external magnetic field is applied \citep[\cf][]{Dend:90}.

It is useful to introduce an order parameter ($\epsilon$) to linearise 
the force equations.  All field quantities are clearly of first order.  
In addition, the change in the velocity of the charged particles is of 
first order
($\delta u^\mu \equiv u^\mu - \overline{u}^\mu \propto \epsilon \exp(iS)$).  
Then, the electromagnetic force upon a single electron is given by
\begin{align}
{\cal F}^\mu_{\mbox{\tiny EM}} =\,& F^{\mu \nu} e u_\nu  \label{lorentz1} \\
=\,& e \overline{u}^\mu \epsilon E^\nu u_\nu
- e \epsilon E^\mu \,\!\overline{u}^\nu u_\nu 
+ e \varepsilon^{\mu\nu\alpha\beta}\,\overline{u}_\alpha\,\epsilon 
B_\beta u_\nu \nonumber
\,.
\end{align}
In the first and third terms only the deviation from $\overline{u}^\mu$
contributes, thus they are of order $\epsilon^2$.  In the second term
$\overline{u}^\mu u_\mu = -1 + {\cal O}(\epsilon)$ hence there
is a first order contribution, and ${\cal F}^\mu_{\mbox{\tiny EM}} = e E^\mu$.
The force is related to $u^\mu$ to first order in $\epsilon$ by 
${\cal F}^\mu_{\mbox{\tiny EM}} = -i \omega m \ \delta\!u^\mu$.  
The current is related to $\delta u^\mu$ by $J^\mu = e n \, \delta u^\mu$.  
Therefore, the conductivity tensor is given by
\begin{equation}
\sigma^\mu_{~\nu} = - \frac{\omega_P^2}{4 \pi i \omega} \delta^\mu_\nu \,,
\label{isohom}
\end{equation}
where $\omega_P \equiv \sqrt{4 \pi e^2 n / m}$ is the plasma frequency.

\subsubsection{Magnetoactive Cold Electron Plasma} \label{MCEP}
In the presence of an externally generated magnetic field, ${\cal B}^\mu$,
(defined in the LFCR frame in the same way as $B^\mu$), the electromagnetic
force upon a single electron is
\begin{align}
{\cal F}^\mu_{\mbox{\tiny EM}} =\,&F^{\mu\nu} e u_\nu \label{lorentz2}\\
=\,&e \overline{u}^\mu \epsilon E^\nu u_\nu
- e \epsilon E^\mu \,\!\overline{u}^\nu u_\nu
+ e \varepsilon^{\mu\nu\alpha\beta}\,\overline{u}_\alpha
\,\left(\epsilon B_\beta + {\cal B}_\beta \right) u_\nu \,.
\nonumber
\end{align}
In contrast to equation (\ref{lorentz1}), there is a first order contribution
from the third term in this case.  Hence, to first order 
${\cal F}^\mu_{\mbox{\tiny EM}} = e E^\mu + e \varepsilon^{\mu\nu\alpha\beta}\,
\overline{u}_\alpha\,{\cal B}_\beta\,u_\nu $.  It is useful to decompose
$\delta\!u^\mu$ and $E^\mu$ into temporal, and spatial components along and
orthogonal to ${\cal B}^\mu$:

\begin{gather}
\delta\!u^\mu_t \equiv \left( \delta\!u_\nu\,\overline{u}^\nu \right)
\overline{u}^\mu 
\;\;,\;\;
\delta\!u^\mu_{\parallel} \equiv \left( 
\frac{{\cal B}^\nu \delta\!u_\nu}{{\cal B}^\alpha{\cal B}_\alpha} \right)
{\cal B}^\mu \,, \nonumber \\
\delta\!u^\mu_{\perp} \equiv \delta\!u^\mu - \delta\!u^\mu_t 
- \delta\!u^\mu_{\parallel} \,,
\end{gather}
\begin{equation}
E^\mu_{\parallel} = \left( 
\frac{{\cal B}^\nu E_\nu}{{\cal B}^\alpha{\cal B}_\alpha} \right) {\cal B}^\mu
\;\;,\;\;
E^\mu_{\perp} = E^\mu - E^\mu_{\parallel} \,.
\end{equation}
With these new definitions it is simple to show that the force equation 
separates into
\begin{align}
-i \omega \delta\!u^\mu_t &= 0 \,, \nonumber \\
-i \omega \delta\!u^\mu_{\parallel} &= \frac{e}{m} E^\mu_{\parallel} \,, 
\\
-i \omega \delta\!u^\mu_{\perp} &= \frac{e}{m} E^\mu_{\perp} + \frac{e}{m} 
\varepsilon^{\mu\nu\alpha\beta}\,\overline{u}_\alpha\,{\cal B}_\beta\,
\delta\!u_{\perp \, \nu} \,. \nonumber
\end{align}
Clearly
$J^\mu_{\parallel} = -\left(\omega_P^2/4\pi i\omega\right) E^\mu_{\parallel}$.
The perpendicular component may be determined by taking a second proper time
derivative whence, to lowest order,
\begin{align}
-\omega^2 \delta\!u^\mu_{\perp} =& -i \omega \frac{e}{m} E^\mu_{\perp} 
\nonumber \\
&+ \frac{e}{m} 
\varepsilon^{\mu\nu\alpha\beta}\,\overline{u}_\alpha\,{\cal B}_\beta 
\left( \frac{e}{m} E_{\perp\,\nu} +\frac{e}{m} 
\varepsilon_{\nu\gamma\sigma\varepsilon}\,\overline{u}^\sigma
{\cal B}^\varepsilon \delta\!u^\gamma_{\perp} \right) \nonumber \\
=& -i \omega \frac{e}{m} E^\mu_{\perp} + \left( \frac{e}{m} \right)^2 
\varepsilon^{\mu\nu\alpha\beta}\,\overline{u}_\alpha\,{\cal B}_\beta\,
E_{\perp \, \nu} \\
&- \left(\frac{e}{m} \right)^2 {\cal B}^\nu {\cal B}_\nu\,
\delta\!u^\mu_{\perp} \,.
\nonumber
\end{align}
Defining $\omega_B^2 \equiv (e/m)^2 {\cal B}^\mu {\cal B}_\mu$ and solving for
$J^\mu_{\perp} = e n \, \delta\!u^\mu_{\perp} $ gives
\begin{equation}
\delta\!u^\mu_{\perp} = \frac{\omega_P^2}{
4\pi\left(\omega_B^2 - \omega^2\right)} \Bigl( 
-i \omega g^{\mu\nu} +
\frac{e}{m} \varepsilon^{\mu\nu\alpha\beta}\,
\overline{u}_\alpha\,{\cal B}_\beta \Bigr) E_{\perp \, \nu} \,.
\end{equation}
After substituting in the expressions for $E^\mu_{\parallel}$ and
$E^\mu_{\perp}$ the total current is given by
\begin{align}
J^\mu &= J^\mu_{\parallel} + J^\mu_{\perp} \nonumber \\
&= -\frac{\omega_P^2}{4 \pi i \omega \left( \omega_B^2 - \omega^2 \right)}
\biggl( -\omega^2 g^{\mu\nu} + \omega_B^2
\frac{{\cal B}^\nu {\cal B}^\mu}{{\cal B}^\alpha{\cal B}_\alpha} \\
&\quad\quad\quad\quad\quad\quad\quad\quad\quad\quad\quad  
- i\omega \frac{e}{m} \varepsilon^{\mu\nu\alpha\beta}\,\overline{u}_\alpha
\,{\cal B}_\beta \biggr) E_\nu \,. \nonumber
\end{align}
As a result, the conductivity tensor can be identified as
\begin{multline}
\sigma_{\mu\nu} = -\frac{\omega_P^2}{4 \pi i \omega \left( \omega_B^2 
- \omega^2 \right)}
\biggl( -\omega^2 g_{\mu\nu} + \omega_B^2  
\frac{{\cal B}_\nu {\cal B}_\mu}{{\cal B}^\alpha{\cal B}_\alpha}
\\
- i\omega \frac{e}{m} \varepsilon_{\mu\nu\alpha\beta}\,\overline{u}^\alpha
\,{\cal B}^\beta \biggr) \,. \label{isohomwB}
\end{multline}
In a flat space, the spatial components of this can be compared to the
standard result \citep[see \eg][]{Boyd-Sand:69,Dend:90}.

\subsection{Ohm's Law for Warm Plasmas} \label{warm}
For AGN and X-ray binaries, accreting plasma near the central compact
object will in general be hot.  Even in low luminosity AGN, accreting
electrons can have $\gamma$'s on the order of $10-10^3$
\citep[see \eg][]{Meli-Falc:01,Nary-Maha-Grin-Poph-Gamm:98}.
In these environments assumption ({\em ii}) in \S\ref{cold}, that the
motions of the electrons are non-relativistic, is no longer valid.

For warm plasmas, ones in which the thermal velocities of the electrons
are significant compared to the phase velocities of the modes, it is
possible to determine the conductivities using the Vlasov equation just
as in flat space \citep[see \eg][]{Dend:90,Boyd-Sand:69,Mont-Tidm:64}:
\begin{equation}
u^\mu \left( \frac{\partial f}{\partial x^\mu} \right)_{p^\mu} 
+
{\cal F}^\mu_{\mbox{\tiny EM}} \left( \frac{\partial f}{\partial p^\mu} \right)_{x^\mu}
= 0 \,,
\label{helmholtz}
\end{equation}
where $p^\mu$ and $f$ are the momentum and distribution function of the
electrons, respectively. The average plasma velocity, $\overline{u}^\mu$,
must now be averaged over temperature in addition to the induced oscillations.
Note that unlike the analyses of warm plasmas in flat space, this must now
be done in a manifestly covariant way. At this point it is necessary to
determine the form of the force, ${\cal F}^\mu_{\mbox{\tiny EM}}$, under
which the system is evolving.

\subsubsection{Isotropic Warm Electron Plasma} \label{IWEP}
In this case ${\cal F}^\mu_{\mbox{\tiny EM}} = F^{\mu\nu} e u_\nu$.
Hence expanding the distribution function in terms of the order parameter
introduced in \S\ref{ICEP} to first order,
$f = f_0 + \epsilon f_1 + {\cal O}(\epsilon^2)$, and inserting into equation
(\ref{helmholtz}) gives
\begin{equation}
u^\mu \left( \frac{\partial f_1}{\partial x^\mu} \right)_{p^\mu} 
+  e F^{\mu\nu} u_\nu \left( \frac{\partial f_0}{\partial p^\mu}
\right)_{x^\mu}
= 0 \,.
\end{equation}
Considering the lowest order in the two length scale expansion of \S\ref{GOA},
this may now be solved for $f_1$:
\begin{equation}
f_1 = \frac{i e u_\nu}{u^\alpha k_\alpha} F^{\mu \nu} 
\left( \frac{\partial f_0}{\partial p^\mu} \right)_{x^\mu} \,,
\end{equation}
which is the covariant analogue of the expressions found in the kinetic theory
literature \citep[see \eg][]{Dend:90}.

Assuming that the plasma was originally charge neutral the current density
is related to the perturbation in the distribution function, $f_1$, by
\begin{equation*}
J^\mu = e \int d^4\!p \, f_1 u^\mu \,.
\end{equation*}  
Then, using equation (\ref{fmunu2a}) this may be written in terms of $E^\mu$ as
\begin{multline}
J^\mu = - \frac{i e^2}{\omega} k^\alpha E^\nu \int d^4\!p 
\frac{u^\mu}{u^\beta k_\beta} \biggl[
u_\alpha \left( \frac{\partial f_0}{\partial p^\nu} \right)_{x^\mu} \\-
u_\nu \left( \frac{\partial f_0}{\partial p^\alpha} \right)_{x^\mu} \biggr] \,.
\end{multline}
From this it is clear that the conductivity tensor is
\begin{multline}
\sigma^\mu_{~\nu} =  -\frac{i e^2}{\omega m} k^\alpha
\int d^4\!p 
\frac{p^\mu}{p^\beta k_\beta} \biggl[ 
p_\alpha \left( \frac{\partial f_0}{\partial p^\nu} \right)_{x^\mu} \\-
p_\nu \left( \frac{\partial f_0}{\partial p^\alpha} \right)_{x^\mu} \biggr] \,.
\label{warmisohom}	
\end{multline}
In order to make a connection with the expression derived in the previous
section it is convenient to integrate this by parts,
\begin{multline}
\sigma_{\mu\nu} =
\frac{i e^2}{\omega m} \int d^4\!p \Bigg[
g_{\mu\nu}
-\frac{k_\mu p_\nu + k_\nu p_\mu}{p_\alpha k^\alpha} \\+ 
\frac{ k_\alpha k^\alpha p_\mu p_\nu}{\left(p_\beta k^\beta \right)^2}
\Bigg] f_0 \,,
\label{warmisohom2}
\end{multline}
where the boundary terms vanish by virtue of the convergence of
$\int d^4\!p \, f_0$.
For the cold plasma, $f_0 = n \delta^4\!(p^\mu - m\overline{u}^\mu)$, thus,
\begin{equation}
\sigma_{\mu\nu} = - \frac{\omega_P^2}{4 \pi i \omega} \left( g_{\mu\nu} + 
\frac{k_\mu \overline{u}_\nu + k_\nu \overline{u}_\mu}{\omega}
+ \frac{ k_\alpha k^\alpha \overline{u}_\mu \overline{u}_\nu}{\omega^2}
\right) \,.
\end{equation}
This differs from the result in \S\ref{ICEP} in two respects: terms
proportional to $\overline{u}_\mu$ and the term proportional to $k_\mu$.
Because the conductivity enters Maxwell's equations only through a contraction
with the electric four-vector, the former are superfluous.  The latter
represents the sonic mode which appears in the kinetic calculation
of the conductivity only in the form of an infinite wavelength mode.
For the two transverse electromagnetic modes ($E^\mu k_\mu = 0$) 
this does agree.

\subsubsection{Magnetoactive Warm Electron Plasma} \label{MWEP}
In the presence of an external magnetic field
${\cal F}^\mu_{\mbox{\tiny EM}}$ has a zeroth order contribution:
\begin{equation}
{\cal F}^\mu_{\mbox{\tiny EM}} = e F^{\mu\nu}\, u_\nu 
+ e F^{\mu\nu}_{\mbox{\tiny Ex}}\, u_\nu \,,
\end{equation}
where, in terms of the external magnetic field (again defined in the LFCR
frame), $F^{\mu\nu}_{\mbox{\tiny Ex}} \equiv 
\varepsilon^{\mu\nu\alpha\beta} \overline{u}_\alpha {\cal B}_\beta $,
(\cf~equation (\ref{fmunua})).
Expanding the Vlasov equation in the perturbation parameter $\epsilon$ 
to first order and in the two length scale expansion (\S\ref{GOA}) now gives,
\begin{equation}
i u^\mu k_\mu f_1 + \frac{e}{m} F^{\mu\nu}_{\mbox{\tiny Ex}}\, p_\nu 
\left( \frac{\partial f_1}{\partial p^\mu} \right)_{x^\mu} = 
- \frac{e}{m} F^{\mu\nu} p_\nu \left( \frac{\partial f_0}{\partial p^\mu} 
\right)_{x^\mu} \,.
\label{liouwarm}	
\end{equation}
At this point it is useful to introduce a function $\eta$ defined implicitly 
by
\begin{equation}
\frac{d}{d \eta} = \frac{e}{m} F^{\mu\nu}_{\mbox{\tiny Ex}}\, p_\nu 
\left( \frac{\partial}{\partial p^\mu} \right)_{x^\mu} \,.
\end{equation}
\citep[\cf][]{Lifs-Pita:81,Kral-Triv:73}.
In terms of $\eta$, the electron momenta are determined by the equation
\begin{equation}
\frac{d p^\mu}{d \eta} = \frac{e}{m} F^{\mu\nu}_{\mbox{\tiny Ex}}\, p_\nu
= \frac{e}{m}  \varepsilon^{\mu\nu\alpha\beta}\,
\overline{u}_\alpha {\cal B}_\beta p_\nu \,.
\label{cyc_eq}
\end{equation}
As in the cold case, this may be reduced to a two dimensional problem by 
an appropriate decomposition of the momentum:
\begin{gather}
p^\mu_t = \left( p_\nu\,\overline{u}^\nu \right)\overline{u}^\mu 
\;\;,\;\;
p^\mu_{\parallel} = \left( 
\frac{{\cal B}^\nu p_\nu}{{\cal B}^\alpha{\cal B}_\alpha} \right) {\cal B}^\mu
\,, \nonumber \\
p^\mu_{\perp} = p^\mu - p^\mu_t - p^\mu_{\parallel} \,,
\end{gather}
In terms of these, the system of equations for $p^\mu$ reduce to
\begin{gather}
\frac{d p^\mu_t}{d \eta} = 0 \,,  \nonumber \\
\frac{d p^\mu_{\parallel}}{d \eta} = 0 \,,  \label{p_des} \\
\frac{d p^\mu_{\perp}}{d \eta} = \frac{e}{m}
\varepsilon^{\mu\nu\alpha\beta}\,
\overline{u}_\alpha {\cal B}_\beta p_{\perp \, \nu} \,. \nonumber
\end{gather}
This last equation is simply that governing cyclotron motion.  Using the 
fact that $d/d\eta$ commutes with the metric (this is because the 
metric depends only upon $x^\mu$ and not $p^\mu$) it may be rewritten as a 
pair of uncoupled, second order ordinary differential equations:
\begin{equation}
\frac{d^2 p^\mu_{\perp}}{d \eta^2} + \omega_B^2 p^\mu_{\perp} = 0 \,.
\label{perpeq}
\end{equation}
This has solutions
\begin{equation}
p^\mu_{\perp} = p_x^\mu \cos(\omega_B \eta + \phi_0)
+ p_y^\mu \sin(\omega_B \eta + \phi_0) \,,
\label{cyc_sol_eq}
\end{equation}
where $p_x^\mu$ and $p_y^\mu$ are a pair of bases which 
span the space perpendicular to $\overline{u}^\mu$ and ${\cal B}^\mu$, 
and $\phi_0$ is a phase factor.  By inserting this solution into equation
(\ref{cyc_eq}) and matching up trigonometric terms, $p_y^\mu$ can be found
in terms of $p_x^\mu$,
\begin{equation}
p^\mu_y = \frac{1}{\omega_B} \varepsilon^{\mu\nu\alpha\beta}
\,\overline{u}_\alpha {\cal B}_\beta
p_{x\,\beta} \,.
\end{equation}
It is possible to now solve for $\eta$ in terms of $p^\mu$, $p_x^\mu$, and
$\phi_0$:
\begin{equation}
\eta = \frac{1}{\omega_B} \left[ \arctan \left( 
\frac{e \varepsilon_{\mu\nu\alpha\beta} p^\mu p_x^\nu
\overline{u}^\alpha {\cal B}^\beta}{m \omega_B p_x^\xi p_\xi} \right) 
- \phi_0 \right] \,.
\label{eta_eq}
\end{equation}

Inserting $p^\mu (\eta)$ into $f_1$ and $f_0$ transform equation 
(\ref{liouwarm}) into a first order differential equation for $f_1$.  This 
has solution
\begin{equation}
f_1 = \left( \mu^{-1} \int \mu \beta_\mu d\eta \right) E^\mu \,,
\end{equation}
where
\begin{gather}
\beta_\mu \equiv \frac{e}{\omega m} k^\nu \left[ p_\nu \left( 
\frac{\partial f_0}{\partial p^\mu} \right)_{x^\mu}
- p_\mu \left( \frac{\partial f_0}{\partial p^\nu} \right)_{x^\mu} \right] 
\,, \\
\mu \equiv \exp \left( - i k_\mu \int \frac{p^\mu}{m} d \eta \right) \,.
\end{gather}
The integral for $\mu$ may be rewritten in terms of $p^\mu$ by using equations 
(\ref{p_des}) and (\ref{perpeq}),
\begin{equation}
\int \left( p^\mu_t + p^\mu_\parallel \right) d\eta = 
\left( p^\mu_t + p^\mu_\parallel \right) \eta \,,
\end{equation}
\begin{align}
\int p^\mu_{\perp} d\eta &= \frac{1}{\omega_B^2} \int 
\frac{d^2 p^\mu_{\perp}}{d \eta^2} = \frac{1}{\omega_B^2} 
\frac{d p^\mu_{\perp}}{d \eta} \nonumber \\
&= \frac{1}{\omega_B^2} \varepsilon^{\mu\nu\alpha\beta}
\,\overline{u}_\alpha {\cal B}_\beta p_{\perp\,\nu} \,.
\end{align}
Thus,
\begin{multline}
\mu = \exp \biggl\{ i \biggl[ \left( \omega \overline{u}_\mu - 
\frac{{\cal B}^\nu k_\nu}{{\cal B}_\alpha {\cal B}^\alpha}
{\cal B}_\mu \right) \eta \\
- \frac{1}{\omega_B^2} \varepsilon_{\mu\nu\alpha\beta}\,k^\nu
\overline{u}^\alpha {\cal B}^\beta \biggr] \frac{p^\mu}{m} \biggr\}\,.
\end{multline}
With equation (\ref{cyc_sol_eq}) this may be treated as a function of $\eta$,
while with equation (\ref{eta_eq}) this may be treated as a function of
$p^\mu$.

As in the previous case, the current four-vector is then found by integrating
over the momentum portion of the phase space.  This gives the conductivity
tensor to be
\begin{equation}
\sigma^\mu_{~\nu} = - \frac{e}{m} \int d^4\!p \, p^\mu \,
\left[\mu^{-1}
\!\!\int \mu \beta_\nu d\eta\right] \left(p^\mu\right) \,, 
\label{warmiso}
\end{equation}
where it has been emphasised that the interior integral is to be treated as
a function of the momenta.

\subsubsection{Conductivity in Quasi-Longitudinal Approximation}
\label{CQLA}
In general, the integrals over $\eta$ in equation (\ref{warmiso}) can be
evaluated in terms of sums of Bessel functions in an analogous fashion to
that typically done for the non-relativistic case
\citep[see \eg][]{Kral-Triv:73}.  Nonetheless, this can
be significantly simplified by considering the case where ({\em i}) $f_0$
is a function of
${\cal P}^2 \equiv p^\mu p_\mu$ and $\epsilon \equiv p^\mu \overline{u}_\mu $
only (typically $f_0$ can be written in the form
$f(\epsilon) \delta({\cal P}^2-m^2)$ where the delta function is required to
place the distribution on the mass-shell), ({\em ii}) 
$\varepsilon^{\mu\nu\alpha\beta}\, \overline{u}_\alpha {\cal B}_\beta k_\mu=0$
(\ie the quasi-longitudinal approximation), ({\em iii}) $\omega_B \ll \omega$,
and ({\em iv}) $f_0$ is such that $p^\mu \overline{u}_\mu /m - 1 \ll 1$
(\ie cool, not hot).

Assumption ({\em i}) simplifies $\beta_\mu$,
\begin{equation}
\beta^\mu = \frac{e}{m\omega}
\frac{\partial f_0}{\partial \epsilon} k_\nu 
\left( \overline{u}^\mu p^\nu - \overline{u}^\nu p^\mu \right) \,.
\label{beta_simp}
\end{equation}
Note that because $\epsilon$ is independent of $\eta$, the terms involving
$f_0$ can now be brought out of the innermost integral in equation
(\ref{warmiso}). Assumption ({\em ii}) gives that $k_\mu p^\mu_{\perp} = 0$
and hence,
\begin{equation}
\mu = {\rm e}^{ i \varpi \eta } \,,
\end{equation}
where $\varpi \equiv k^\mu p_\mu / m$.
Therefore, the two integrals that must be done are
\begin{equation}
\int p_{\parallel}^{\mu} {\rm e}^{ i \varpi \eta } d \eta
=
p_{\parallel}^{\mu} \frac{ \mu } {i \varpi } \,,
\end{equation}
and
\begin{multline}
\int p_{\perp}^{\mu} {\rm e}^{ i \varpi \eta } d \eta
=\\
\left(
g^{\mu \nu}
- \frac{e}{i \varpi m}
\varepsilon^{\mu\nu\alpha\beta}\,
\overline{u}_\alpha {\cal B}_\beta
\right)
p_{\perp \, \nu}
\frac{ \varpi^2 }
{ \varpi^2 - \omega_B^2 } 
\frac{ \mu } {i \varpi } \,.
\end{multline}
Therefore, in the quasi-longitudinal approximation,
\begin{multline}
f_1 = \frac{e}{i \varpi m} \frac{\partial f_0}{\partial \epsilon}
\frac{1}{\varpi^2-\omega_B^2}
\bigg[ \varpi^2 g_{\mu \nu}
- \omega_B^2 \frac{{\cal B}_\mu {\cal B}_\nu}{{\cal B}^\alpha {\cal B}_\alpha}
\\
+ \frac{i \varpi e}{m}
\varepsilon_{\mu\nu\alpha\beta} \overline{u}^\alpha {\cal B}^\beta
\bigg] p^\nu E^\mu \,,
\end{multline}
where the definitions of $p_{\parallel}^\mu$, $p_{\perp}^\mu$, and $E^\mu$
were used.  In the quasi-longitudinal approximation, $E^\mu$ is
orthogonal to the external magnetic field, ${\cal B}^\mu$.  As a result,
the there are only two integrals that must be done in order to find the
conductivity tensor:
\begin{align}
I_1^{\mu\nu} &= -\frac{i \omega}{m} \int d^4\!p
\frac{i\varpi}{\varpi^2 - \omega_B^2}
p^\mu p^\nu \frac{\partial f_0}{\partial \epsilon} \nonumber \\
I_2^{\mu\nu} &= -\frac{i \omega}{m} \int d^4\!p
\frac{1}{\varpi^2 - \omega_B^2} p^\mu p^\nu
\frac{\partial f_0}{\partial \epsilon} \,.
\label{Iints}
\end{align}
In terms of these, the conductivity is
\begin{equation}
\sigma^\mu_{~\nu} = - \frac{e^2}{i \omega m} \left(
I_1^{\mu\gamma} g_{\gamma\nu}
- \frac{e}{m}
I_2^{\mu\gamma} \varepsilon_{\nu\gamma\alpha\beta} \overline{u}^\alpha
{\cal B}^\beta
\right) \,.
\label{cond_wqla}
\end{equation}
From equation (\ref{beta_simp}) it follows that
\begin{multline}
p_\mu p_\nu \frac{\partial f_0}{\partial \epsilon}
=
\frac{ p_\mu k^\alpha}{\omega} \left(
p_\alpha \frac{\partial f_0}{\partial p^\nu}
-
p_\nu \frac{\partial f_0}{\partial p^\alpha}
\right) \\
-
\frac{ k^\alpha p_\alpha}{\omega} p_\mu \overline{u}_\nu
\frac{\partial f_0}{\partial \epsilon} \,.
\end{multline}
Noting that the $I^{\mu\nu}$ will only be contracted on the second index with
terms orthogonal to $\overline{u}^\mu$ (for $I_1^{\mu\nu}$ this is the
electric field), the $I^{\mu\nu}$ are given by,
\begin{align}
I_1^{\mu\nu} &= -i \int d^4\!p
\frac{i\varpi}{\varpi^2 - \omega_B^2}
p^\mu \left( \varpi g^{\nu \alpha} - \frac{p^\nu k^\alpha}{m} \right)
\frac{\partial f_0}{\partial p^\alpha} \nonumber \\
I_2^{\mu\nu} &= -\frac{i}{m} \int d^4\!p
\frac{1}{\varpi^2 - \omega_B^2} p^\mu
\left( \varpi g^{\nu \alpha} - p^\nu k^\alpha \right)
\frac{\partial f_0}{\partial p^\alpha} \,.
\label{Iints2}
\end{align}

Because there is already a term linear in $\omega_B$ in equation
(\ref{cond_wqla}), to lowest order in assumption ({\em iii}) $\omega_B^2$
may be neglected in the $I^{\mu\nu}$.  Thus,
\begin{align}
I_1^{\mu\nu} &= \int d^4\!p \,
p^\mu \left( g^{\nu \alpha} - \frac{p^\nu k^\alpha}{m \varpi} \right)
\frac{\partial f_0}{\partial p^\alpha} \nonumber \\
I_2^{\mu\nu} &= -i \int d^4\!p \,
\frac{p^\mu}{\varpi}
\left( g^{\nu \alpha} - \frac{p^\nu k^\alpha}{m \varpi} \right)
\frac{\partial f_0}{\partial p^\alpha} \,.
\label{Iints_app1}
\end{align}
These may be integrated by parts to produce
\begin{align}
I_1^{\mu\nu} &= - \int d^4\!p \, f_0
\left(
g^{\mu\nu} - \frac{p^\mu k^\nu + p^\nu k^\mu}{m \varpi}
+ \frac{p^\mu p^\nu}{m^2 \varpi^2} k^\alpha k_\alpha
\right) \nonumber \\
I_2^{\mu\nu} &= i \int d^4\!p \, \frac{f_0}{\varpi}
\left(
g^{\mu\nu} - \frac{2 p^\mu k^\nu + p^\nu k^\mu}{m \varpi}
+ 2 \frac{p^\mu p^\nu}{m^2 \varpi^2} k^\alpha k_\alpha
\right) \,.
\end{align}
Note that in this case, $I_1^{\mu\nu}$ is simply the integral that had to be
done for the warm isotropic plasma (\cf equation (\ref{warmisohom2})).

Assumption ({\em iv}) enters by expanding $\varpi$ about $\omega$.  Define
$\wp^2 \equiv \epsilon^2 - m^2$, \ie~$\wp$ is the magnitude of the spatial
components of the momentum in the LFCR frame.  Then, to second order in $\wp$,
\begin{multline}
\varpi^j \simeq (-\omega)^j \left[ 1
- j \left(\frac{p^\mu {\cal B}_\mu k^\nu {\cal B}_\nu}{m \omega
{\cal B}^\alpha {\cal B}_\alpha} \right) \right. \\
\left.
+ \frac{j(j-1)}{2} \left(\frac{p^\mu {\cal B}_\mu k^\nu {\cal B}_\nu}{m \omega
{\cal B}^\alpha {\cal B}_\alpha} \right)^2
+ j \frac{\wp^2}{2 m^2}
\right] \,.
\end{multline}
Thus,
\begin{align}
I_1^{\mu\nu} &\simeq - \int d^4\!p \, f_0
\left( g^{\mu\nu} + \frac{p^\mu p^\nu}{m^2}
\frac{k^\alpha k_\alpha}{\omega^2} \right) \nonumber \\
I_2^{\mu\nu} &\simeq - i \int d^4\!p \, \frac{f_0}{\omega}
\Bigg\{
\left[
1
- \frac{\wp^2}{2 m^2}
+ \left(\frac{p^\mu {\cal B}_\mu k^\nu {\cal B}_\nu}{m \omega
{\cal B}^\alpha {\cal B}_\alpha} \right)^2
\right] g^{\mu\nu} \nonumber \\
&\quad\quad\quad\quad\quad\quad\quad\quad\quad\quad\quad\quad\quad
+ 2 \frac{p^\mu p^\nu}{m^2} \frac{k^\alpha k_\alpha}{\omega^2}
\Bigg\} \,,
\end{align}
where terms odd in $p^\mu$ and terms $\propto k^\nu$ have been dropped.
The former is due to the fact that $f_0$ has been chosen to be an isotropic
function of the spatial components of the momentum in the LFCR frame and
hence any odd terms will vanish upon integration.  The latter is allowed
because, as stated earlier, these will only have significance when
contracted with terms orthogonal to $k^\mu$ (for $I_2^{\mu\nu}$ this is
results from the quasi-longitudinal approximation in which $k^\mu$ can
be written in terms of $\overline{u}^\mu$ and ${\cal B}^\mu$ only).
From symmetry it is clear that
\begin{equation}
\int d^4\!p \, f_0
\frac{ \left( p^\mu {\cal B}_\mu \right)^2}{{\cal B}^\alpha {\cal B}_\alpha}
= \frac13 n m^2 \left< f_0 \right>_2 \,,
\end{equation}
where
\begin{equation}
\left< f_0 \right>_2 \equiv \frac{1}{n m^2} \int d^4\!p \, f_0 \wp^2 \,.
\end{equation}
In addition, the off-diagonal components of the integrals over $p^\mu p^\nu$
will vanish due to the symmetry of $f_0$.  Because adding terms
$\propto \overline{u}^\nu$ will not alter the physical solutions, it is
possible to replace $\int d^4\!p \, p^\mu p^\nu f_0$ with
$\frac13 n m^2 \left< f_0 \right>_2 g^{\mu\nu}$.  Lastly, note that
\begin{equation}
\frac{ \left( k^\nu {\cal B}_\nu \right)^2}{{\cal B}^\alpha {\cal B}_\alpha}
= \omega^2 + k^\alpha k_\alpha \,.
\end{equation}
Therefore, the
$I^{\mu\nu}$ are given by
\begin{equation}
I_1^{\mu\nu} \simeq n {\cal I}_1 g^{\mu\nu}
\quad \mbox{and} \quad
I_2^{\mu\nu} \simeq \frac{n}{i \omega} {\cal I}_2 g^{\mu\nu} \,,
\end{equation}
where
\begin{align}
{\cal I}_1 &\equiv 1
+ \frac13 \frac{k^\alpha k_\alpha}{\omega^2} \left< f_0 \right>_2
\nonumber \\
{\cal I}_2 &\equiv 1
- \frac16 \left< f_0 \right>_2
+ \frac{k^\alpha k_\alpha}{\omega^2}  \left< f_0 \right>_2 \,.
\end{align}
Because the terms multiplying ${\cal I}_2$ in the conductivity are already of
first order (the order of $\omega_B$ is necessarily equal to or smaller than
that of $\wp$ for the approximations thus far to hold), to second order in
small quantities in the conductivity, ${\cal I}_2 \simeq 1$. 
As a result, with the lowest order finite temperature corrections the
conductivity is given by
\begin{equation}
\sigma_{\mu\nu} \simeq - \frac{\omega_P^2}{4\pi i \omega} \left(
{\cal I}_1 g_{\mu\nu}
- \frac{e}{i\omega m}
\varepsilon_{\mu\nu\alpha\beta} \overline{u}^\alpha {\cal B}^\beta 
\right)
\end{equation}
For the cold plasma ${\cal I}_1 = 1$ and this does reduce to the
appropriate expansion of the conductivity derived in \S\ref{MCEP}.

\subsection{Dispersion Relations}
Given the conductivities derived in \S\ref{cold} \& \S\ref{warm} it is now
possible to obtain the associated dispersion relations.  It is instructive to
compare these to the dispersion relation for massive particles (de Broglie
waves):
\begin{equation}
D(k_\mu,x^\mu) = k^\mu k_\mu + m^2 \,.
\label{deBroglie}
\end{equation}
That this does produce the time-like geodesics when inserted into the ray
equations is demonstrated in appendix \ref{massive_particles}.

\subsubsection{Isotropic Electron Plasma}
\label{IEP}
The conductivity tensor obtained in \S\ref{ICEP} for the isotropic cold
electron plasma yields the dispersion tensor
\begin{equation}
\Omega^\mu_{~\nu}
= \left( k^\alpha k_\alpha + \omega_P^2 \right) \delta^\mu_\nu
- k^\mu k_\nu \,.
\end{equation}
For the transverse modes, this gives the dispersion relation
\begin{equation}
D(k_\mu,x^\mu) = k^\mu k_\mu + \omega_P^2 \,,
\label{disp_iso}
\end{equation}
\citep[\cf][]{Kuls-Loeb:92}.
For constant density plasmas this is nothing more than the massive particle
equation, \cf~equation (\ref{deBroglie}).  For plasmas with spatially
varying densities this leads to a variable effective ``mass''.  Hence in
general, photons in plasmas will not follow geodesics.  This is a
representation of the refractive nature of the plasma.

\subsubsection{Quasi-Longitudinal Approximation for the Cold Electron Plasma}
\label{QLA}
When magnetic fields are present it is necessary to utilise the conductivity
tensor obtained in \S\ref{MCEP}.
In the quasi-longitudinal approximation the wave four-vector is parallel
to the external magnetic field.  In this approximation, the modes are
transverse.
This follows from the fact that in the LFCR frame this is true and that
since this is a local property expressible in covariant form, it must also
be true in an arbitrary frame.  This can be explicitly verified by comparison
with the results of \S\ref{GMP} where the general case is considered.

Under these conditions the dispersion tensor takes the form
\begin{equation}
\Omega^\mu_{~\nu} = \alpha \delta^\mu_\nu - i \gamma M^\mu_{~\nu} \,,
\label{disp_eq_ql}
\end{equation}
where $\alpha$, $\gamma$, and $M_{\mu\nu}$ are defined by
\begin{gather}
\alpha \equiv k^\mu k_\mu - \delta \omega^2
\,,\quad
\gamma \equiv \delta \omega \left(\frac{e}{m}\right) \,,
\label{coeffs} \\
\delta \equiv \frac{\omega_P^2}{\omega_B^2 - \omega^2}
\,,\quad M_{\mu\nu}=-M_{\nu\mu} \equiv
\varepsilon_{\mu\nu\alpha\beta}\,\overline{u}^\alpha {\cal B}^\beta 
\,. \nonumber
\end{gather}
Taking the determinant of $\Omega^\mu_{~\nu}$ yields
\begin{align}
\det \Omega^\mu_{~\nu}
&= \alpha^4 - \alpha^2 \gamma^2 {\cal B}^\mu {\cal B}_\mu \nonumber \\
&= \alpha^2 \left( \alpha - \delta \omega \omega_B \right)
\left( \alpha + \delta \omega \omega_B \right) = 0 \,.
\end{align}
The two modes corresponding to $\alpha=0$ are the sonic mode and the 
unphysical mode proportional to $\overline{u}^\mu$ which is eliminated by
the condition that $\overline{u}_\mu E^\mu = 0$.  The other two modes have
dispersion relations
\begin{align}
D\left( k_\mu, x^\mu \right)
&= \alpha \pm \delta \omega \omega_B \nonumber \\
&= k^\mu k_\mu + \frac{\omega \omega_P^2}{\omega \pm \omega_B} \,.
\label{disp_ql}
\end{align}
As with equation (\ref{disp_iso}), this dispersion relation also has
a term that could be identified with the mass in equation (\ref{deBroglie}).
In contrast with equation (\ref{disp_iso}), now that ``mass'' depends upon
the polarisation eigenmode.  As a result, different eigenmodes will propagate
differently.  Again this is an expression of the dispersive nature of a
magnetised plasma.

In addition to dispersion, a noticeable departure from its non-relativistic
analogue is the presence of $k_\mu$ in the definition of $\omega$.  This is not
surprising since it is the most general Lorentz covariant extension of the
quasi-longitudinal dispersion relation.  Of interest is the fact that the
dispersion relation is now cubic in the magnitude of $\vec{k}$, $\kappa$.
Because two roots clearly exist in the low density limit, a third root must
also exist.  This results in a new branch in the dispersion relation.  This
will be explored in more detail in \S\ref{BPF}.

\subsubsection{Quasi-Longitudinal Approximation for the Warm Electron Plasma}
\label{QLA_WP}
For the conductivity derived in \S\ref{CQLA}, this is identical to the
previous section, where $\alpha$ and $\delta$, are replaced by
$k^\mu k_\mu + {\cal I}_1 \omega_P^2$ and $-\omega_P^2/\omega^2$.
Then,
\begin{align}
D\left( k_\mu, x^\mu \right)
&= \alpha \pm \omega_P^2 \frac{\omega_B}{\omega} \nonumber \\
&= k^\mu k_\mu + {\cal I}_1 \omega_P^2
\pm \omega_P^2 \frac{\omega_B}{\omega} \nonumber \\
&= \left( 1 + \frac13 \frac{\omega_P^2}{\omega^2} \left< f_0 \right>_2 \right)
k^\mu k_\mu + \omega_P^2 \pm \omega_P^2 \frac{\omega_B}{\omega} \,.
\label{disp_wql}
\end{align}
For a thermal electron distribution, $\left< f_0 \right>_2 = 3 kT / m$ and
hence
\begin{equation}
\frac13 \frac{\omega_P^2}{\omega^2} \left< f_0 \right>_2
= \frac{\omega_T^2}{\omega^2}
\quad \mbox{where} \quad
\omega_T^2 = \frac{k T}{m} \omega_P^2 \,.
\end{equation}
Note that $\omega_T$ is related to the Debye frequency, $\omega_D$, by
$\omega_T = \omega_P^2/\omega_D$.  Thus, including the lowest
order finite temperature corrections, the dispersion relation in the
quasi-longitudinal approximation is
\begin{equation}
D\left( k_\mu, x^\mu \right)
=
\left( 1 + \frac{\omega_T^2}{\omega^2} \right) k^\mu k_\mu
+ \omega_P^2 \pm \omega_P^2 \frac{\omega_B}{\omega} \,.
\end{equation}

\subsubsection{General Magnetoactive Cold Pair Plasma}
\label{GM_CPP}
The conductivity for the pair plasma may be obtained by adding the
conductivities for the electrons and the positrons,
\begin{align}
\sigma_{\mu\nu}^{\mbox{\tiny pair}}
&= \sigma_{\mu\nu}^{e^-} + \sigma_{\mu\nu}^{e^+} \nonumber \\
&= -\frac{\omega_P^2}{4 \pi i \omega \left( \omega_B^2 
- \omega^2 \right)}
\left(
-\omega^2 g_{\mu\nu} + \omega_B^2  
\frac{{\cal B}_\nu {\cal B}_\mu}{{\cal B}^\alpha{\cal B}_\alpha}
\right) \,,
\end{align}
where now the plasma frequency is defined in terms of the sum of the
number densities of the electrons and positrons.  The resulting dispersion
tensor is
\begin{equation}
\Omega_{\mu\nu} = \alpha g_{\mu\nu} - k_\mu k_\nu
+ \beta {\cal B}_\mu {\cal B}_\nu \,,
\label{disp_tens_pp}
\end{equation}
where $\alpha$, $\gamma$, $\delta$, and $M_{\mu\nu}$ are defined as in
equation (\ref{coeffs}), and $\beta\equiv\delta (e/m)^2$.  In addition to
the requirement that $\Omega^\mu_{~\nu} E^\nu = 0$, $E^\mu$ must be orthogonal
to $\overline{u}^\mu$.  As a result, it is necessary to alter
$\Omega^\mu_{~\nu}$ in such a way that it explicitly separates the eigenmodes
orthogonal to $\overline{u}^\mu$ from the unphysical mode.  This can be
trivially accomplished by adding a term $-\omega k_\mu \overline{u}_\nu$ to the
dispersion tensor.  Note that this does not change the dispersion equation
for the physical modes because $E^\mu \overline{u}_\mu = 0$.  Thus, consider
\begin{equation}
\Omega_{\mu\nu} = \alpha g_{\mu\nu}
- k_\mu \left( k_\nu - \omega \overline{u}_\nu \right)
+ \beta {\cal B}_\mu {\cal B}_\nu \,,
\end{equation}
instead of the dispersion tensor given in equation (\ref{disp_tens_pp}).
For this dispersion tensor, the unphysical mode is trivially found to be
$\overline{u}^\mu$, with dispersion relation $D=\alpha$.  As in \S\ref{QLA}
the dispersion relations can be found by taking the determinant of the
dispersion tensor:
\begin{multline}
\det \Omega^\mu_{~\nu} = -(1+\delta)\omega^2 \alpha^2
\Biggl[ \alpha + \delta\omega_B^2 \\
- \frac{\delta}{1+\delta} \left( \frac{e {\cal B}^\mu k_\mu}{m \omega}
\right)^2
\Biggr] \,,
\end{multline}
where the definition of $\alpha$ was used.  Therefore, the dispersion
relations for the two electromagnetic modes are
\begin{align}
D_1\left( k_\mu, x^\mu \right) &=
k^\mu k_\mu - \frac{\omega_P^2}{\omega_B^2-\omega^2} \\
D_2\left( k_\mu, x^\mu \right) &=
k^\mu k_\mu + \omega_P^2
- \frac{\omega_P^2}{\omega_P^2 + \omega_B^2 - \omega^2}
\left( \frac{e {\cal B}^\mu k_\mu}{m \omega} \right)^2 \nonumber \,.
\end{align}
It is straightforward to show that $D_1$ and $D_2$ correspond to the
extraordinary and ordinary modes, respectively, by considering the
transverse limit (${\cal B}^\mu k_\mu = 0$).

\subsubsection{General Magnetoactive Cold Electron Plasma} \label{GMP}
For the general case, no approximations, except those used to derive equations
(\ref{disp_eq}) and (\ref{isohomwB}), are made.  In this case, inserting the 
conductivity tensor obtained in \S\ref{MCEP} into equation (\ref{disp_tensor})
gives
\begin{multline}
\Omega_{\mu\nu} = 
k^\alpha k_\alpha g_{\mu\nu} - k_\mu k_\nu
- \frac{\omega_P^2}{\left( \omega_B^2 - \omega^2 \right)}
\biggl( \omega^2 g_{\mu\nu} \\ 
- \omega_B^2 \frac{{\cal B}_\mu {\cal B}_\nu}{{\cal B}^\alpha{\cal B}_\alpha}
+ i\omega \frac{e}{m} \varepsilon_{\mu\nu\alpha\beta}\,\overline{u}^\alpha
\,{\cal B}^\beta \biggr) \,. 
\end{multline}
Collecting the coefficients of like tensors gives
\begin{equation}
\Omega^\mu_{~\nu} =
\alpha \delta^\mu_{~\nu} - k^\mu k_\nu + \beta {\cal B}^\mu {\cal B}_\nu
- i \gamma M^\mu_{~\nu} \,,
\end{equation}
where $\alpha$, $\beta$, $\gamma$, $\delta$, and $M_{\mu\nu}$ are defined as in
\S\ref{QLA} and \S\ref{GM_CPP}.  As in the previous section, it is useful
to add a term proportional to $\overline{u}_\nu E^\nu$ to the dispersion
equation.  Hence consider
\begin{equation}
\Omega^\mu_{~\nu} =
\alpha \delta^\mu_\nu - k^\mu \left( k_\nu 
- \omega \overline{u}_\nu \right) + \beta {\cal B}^\mu {\cal B}_\nu
- i \gamma M^\mu_{~\nu} \,.
\end{equation} 
Proceeding as in the previous sections, the scalar dispersion relations
corresponding to the different eigenmodes can be found by considering
the determinant of the dispersion tensor:
\begin{multline}
\det \Omega^\mu_{~\nu}
=
\alpha \biggl\{ \alpha^3 + \left[ \beta {\cal B}^\mu {\cal B}_\mu
- \left( k^\mu k_\mu + \omega^2 \right) \right] \alpha^2 \\
- \left[ \delta \omega_B^2 \left( k^\mu k_\mu + \omega^2 \right)
- \delta \left( \frac{e}{m} {\cal B}^\mu k_\mu \right)^2
+ \delta^2 \omega^2 \omega_B^2 \right] \alpha \\
- \delta^2 \omega^2 \left[ \delta \omega_B^4 
- \left( \frac{e}{m} {\cal B}^\mu k_\mu \right)^2 \right] \biggr\} \,.
\end{multline}
Inserting the definition of $\alpha$ reduces the terms in the braces to a
quadratic in $k^\mu k_\mu$, which may be solved to produce the desired
dispersion relation:
\begin{multline}
D\left( k_\mu, x^\mu \right)
= \\
k^\mu k_\mu - \delta \omega^2
- \frac{\delta}{2 \left( 1 + \delta \right)}
\Biggl\{ \Biggl[ \left( \frac{e {\cal B}^\mu k_\mu}{m \omega} \right)^2
- \left( 1 + 2 \delta \right) \omega_B^2 \Biggr] \\
\pm 
\sqrt{
\left( \frac{e {\cal B}^\mu k_\mu}{m \omega} \right)^4
+ 2 \left( 2 \omega^2 - \omega_B^2 - \omega_P^2 \right)
\left( \frac{e {\cal B}^\mu k_\mu}{m \omega} \right)^2
+ \omega_B^4
}
\Biggr\} \,.
\label{disp_gen}
\end{multline}
This is a covariant extension of the Appleton--Hartree dispersion relation
\citep[see \eg][]{Boyd-Sand:69}.  As in the previous
two sections, this continues to bear a resemblance to the dispersion
relation for massive particles.  Again the effective ``mass'' depends upon
position and the polarisation eigenmode.  Additionally, it now depends upon
the direction of propagation relative to the external magnetic field as well.

\section{Example Applications} \label{Examples}

In \S\ref{Theory} the general theory of a covariant magnetoionic theory was
presented for electron-ion (in the Appelton-Hartree limit) and pair plasmas.
While astrophysical plasmas will in general be
warm, the cold electron plasma does provide a instructive setting in which
to highlight some of the similarities and differences that a fully general
relativistic magnetoionic theory has compared to general relativity or
plasma effects alone.

\subsection{Bulk Plasma Flows} \label{BPF}
\begin{figure}
\begin{center}
\includegraphics[width=\columnwidth]{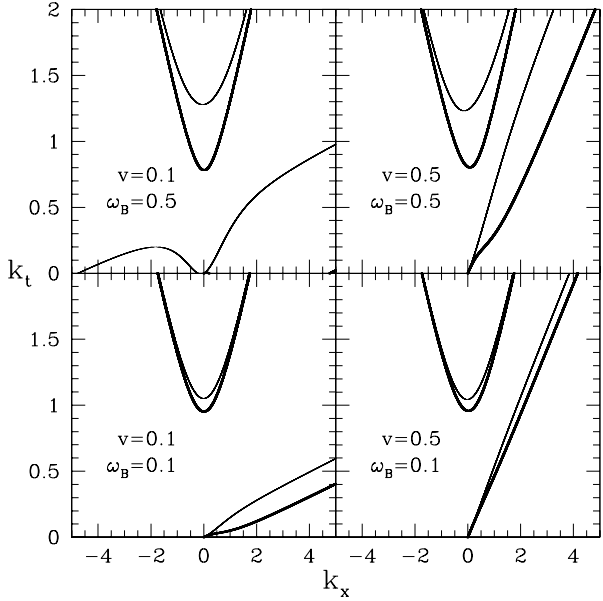}
\end{center}
\caption{The dispersion diagram at a number of magnetic field strengths
and velocities for a relativistic bulk plasma flow.  The frequency scale
is set by $\omega_P=1$.  The ordinary (extraordinary) eigenmode is shown
by the thick (thin) line. Note that the dispersion diagrams are asymmetric
due to the plasma motion.}
\label{qla_bpf_fig}
\end{figure}

A number of effects will appear in special relativistic plasma flows.
The covariant formulation of magnetoionic theory can have implications for
the structure of the dispersion
relation.  As briefly mentioned in \S\ref{QLA}, the equation for the magnitude
of the spatial part of the wave vector is now cubic.  This is essentially due
to Doppler shifting.  Thus these effects should appear in relativistic bulk
plasma flows as well as in regions of strong frame dragging (\eg near the 
ergosphere of a Kerr hole).

For a relativistic bulk flow (in the $x$ direction)
\begin{equation}
\omega = \frac{k_t - v k_x \cos\theta }{ 
\sqrt{1-v^2} } \,,
\label{bpf_omega}
\end{equation}
where $\theta$ is
the angle between the wave vector and the motion and $v$ is the velocity
of the motion.  Clearly the coupling between the previously mentioned third
branch depends upon both $v$ and $\theta$, being strongest when $\theta=0$.
Shown in figure \ref{qla_bpf_fig} are the
quasi-longitudinal dispersion relations for a relativistic bulk flow for
a number of velocities and magnetic field strengths and $\theta=0$.
The frequencies are measured in units of the plasma frequency, making this
otherwise scale invariant.  Note that a whistler-like branch appears for
the ordinary mode which is not present in the non-relativistic theories.
Similar to the whistler branch of the extraordinary mode, it is asymmetric
due to the bulk motion.  In the limit of vanishing plasma density this branch
does not transform into a vacuum branch, in much the same manner as portions
of the whistler.  Therefore, this mode cannot escape from the plasma,
necessarily reflecting at the surfaces of the plasma distribution.  This
may have implications for the pressure balance in thick disks with large
velocity shears and jets, even at frequencies where these are optically thin.

In bulk plasma flows the new branch appears because the velocity mixes
the spatial and temporal components of $k^\mu$.  In a Kerr spacetime, frame
dragging is responsible for mixing these components.  In this case
\begin{equation}
\omega = \sqrt{-g^{tt}} \left( k_t + \frac{g^{\phi t}}{g^{tt}} k_\phi \right)
\,.
\end{equation}
This is similar to equation (\ref{bpf_omega}) with the role of the
velocity being taken by $g^{\phi t}/g^{tt}$.  Hence, the overall effect
is qualitatively the same; a new branch similar to the whistler appears
for the ordinary mode.

\subsection{Isotropic Plasmas and Particle Dynamics}
\label{IPPD}
In both special and general relativistic settings, the propagation of photons
through an isotropic (field free) plasma can be represented in a manner
analogous to that of particle dynamics in a potential
\citep[see \eg][for the non-relativistic case]{Thom-Blan-Phin:94}.
Following the
manipulations in appendix \ref{massive_particles}, it is straight forward
to show that for the dispersion relation given in \S\ref{IEP},
$D = k^\mu k_\mu + \omega_P^2$, that
\begin{equation}
v^\nu \nabla_\nu v^\mu = - \nabla^\mu 4 \omega_P^2 \,,
\label{p_dynamics}
\end{equation}
where $v^\mu \equiv d x^\mu / d \tau$,
\ie~$4 \omega_P^2$ acts as a potential in which the the photons propagate
(the factor of $4$ is due to the particular affine parameter chosen, namely
that associated with the choice of the dispersion relation given above).

For plasmas in which magnetoionic effects are not significant to the photon
propagation (magnetoionic effects may still be important for emission and
the propagation of polarisation) this allows a somewhat more simplified
analysis.  If enough symmetries are present, then the rays may be determined
via direct integration.  For example, consider a stationary, spherically
symmetric plasma distribution around a Schwarzschild black hole.  In this case
equation (\ref{p_dynamics}) shows that $v_t$ and $v_\phi$ are conserved,
associated with the time and azimuthal Killing vector fields, respectively.
Therefore, with the dispersion relation,
\begin{align}
\frac{d t}{d \tau} &= v^t
= g^{tt} v_t = -\left(1-\frac{2M}{r}\right)^{-1} v_t \nonumber \\
\frac{d \phi}{d \tau} &= v^\phi
= g^{\phi \phi} v_\phi = \frac{v_\phi}{r^2} \nonumber \\
\frac{d r}{d \tau} &= v^r
= \sqrt{v_t^2 - \left(1-\frac{2M}{r}\right)
\left( \frac{L}{r^2} + 4 \omega_p^2 \right)} \,,
\end{align}
Which may be directly integrated to give the ray as a function of the
affine parameter $\tau$ in precisely the same fashion as is typically done
to find the particle orbits of the Schwarzschild metric.

\subsection{Photon Capture Cross Sections} \label{PCCS}
\begin{figure}
\begin{center}
\includegraphics[width=\columnwidth]{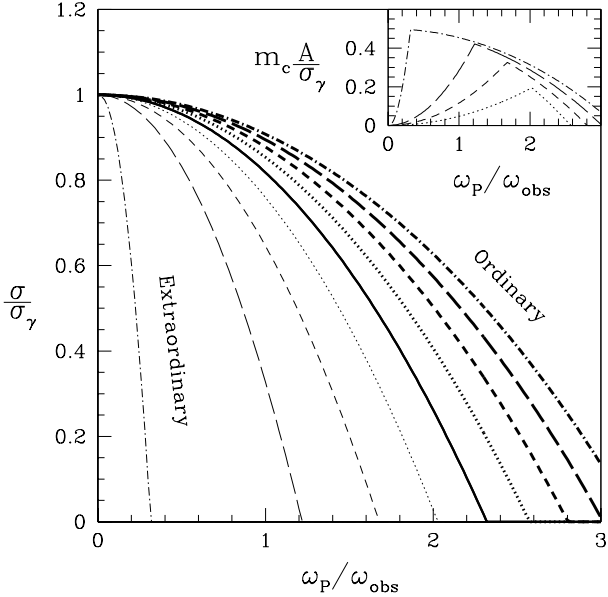}
\end{center}
\caption{Photon capture cross sections in units of the vacuum capture
cross section, $\sigma_{\gamma} = 27 \pi M^2$, for the quasi-longitudinal
approximation as a function of plasma density
($\omega_P/\omega_{\mbox{\tiny obs}}$ is the value of the plasma frequency at
$r=3M$) at a number of magnetic field strengths.  The solid, dotted, short
dashed, long dashed, and dash-dotted lines correspond to 
$\omega_B/\omega_{\mbox{\tiny obs}} = 0,\,0.7,\,1.4,\,2.1,\,
\mbox{and }2.8$, respectively, at $r=3M$.  The inset shows the circular
polarisation fraction, $m_c$, in terms of the effective emission area, $A$,
for the same set of magnetic field strengths.}
\label{cross_sections}
\end{figure}

In the vicinity of a black hole, polarisation can arise even in the case of
a gray emissivity.  This occurs when
one mode is preferentially captured by the black hole due to dispersive
plasma effects.  Even without a method for performing the radiative transfer,
this can be estimated by considering the photon capture cross section of
Schwarzschild black hole.  It is necessary to provide a plasma geometry
-- the plasma density, velocity, and magnetic field -- as functions of
position.  Here, the density is given by the self-similar Bondi solution,
$\omega_P \propto r^{-3/4}$.  The magnetic field is chosen to be a fixed
fraction of the equipartion value, $\omega_B \propto r^{-5/4}$.  Finally,
the velocity is chosen such that the plasma has zero angular momentum,
\ie~$\overline{u}_t = 1/\sqrt{-g^{tt}}$ and
$\overline{u}_r=\overline{u}_\theta=\overline{u}_\phi=0$.  While this 
doesn't correspond to a realistic accretion flow, it does provide
insight into the type of effects dispersion can have. In order
to further simplify the problem the quasi-longitudinal approximation was used.
Typically this is a good approximation, only failing when the angle between
$k^\mu$ and ${\cal B}^\mu$ is within $\sim \omega_P^2 \omega_B /\omega^3$ of
$\pi/2$.  This dispersive polarisation mechanism produces primarily
circular polarisation for the same reason.

Shown in figure \ref{cross_sections} are these cross sections for a number
of different plasma densities (through $\omega_P$) and magnetic field
strengths (through $\omega_B$).  These are both scaled by the observed
frequency at infinity, and hence are not tied to any particular frequency
scale.  The capture cross section of the extraordinary mode decreases
more rapidly than that of the ordinary mode, with increasing density.  The
disparity between the two capture cross sections increases with increasing
magnetic field strength.

This can be a very
efficient manner of creating polarisation over the inner portions of the
accretion flow.  However, far from the hole (outside the inner $5-10 M$) this
becomes a small effect.  As a result, the fraction of polarisation produced
depends upon the magnitude of the diluting emission from regions of the
accretion flow distant from the hole.  Nonetheless, it is possible to
parameterise the unknown emission in terms of an effective emitting area
(the details of which still depend upon the details of the accretion flow).
Shown in the inset of figure \ref{cross_sections} is the circular polarisation
fraction scaled by the effective emission area in units of the vacuum
photon capture cross section.

\subsection{Tracing Rays} \label{TR}
\begin{figure}
\begin{center}
\includegraphics[width=\columnwidth]{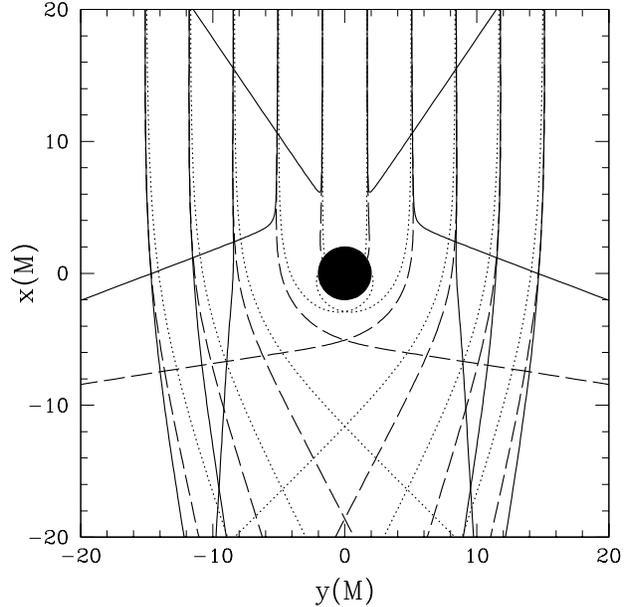}
\end{center}
\caption{ The paths of the ordinary and extraordinary polarisation eigenmodes
in the vicinity of a Schwarzschild black hole are shown by the dashed and solid
lines, respectively, for a number of impact parameters.  The dotted lines show
the null geodesics for comparison.  The $x$ axis lies along the ray paths at
infinity, and the $y$ axis is orthogonal to both, the $x$ axis and the slice
of impact parameters considered.  The plasma density
is $\propto r^{-3/2}$ and $\omega_P (r=3 M) = \omega_{\mbox{\tiny obs}}$.
The magnetic field has a split monopole geometry with its strength
$\propto r^{-5/4}$ and $\omega_B (r=3 M) = 2 \omega_{\mbox{\tiny obs}}$.
The horizon is shown by the filled region in the centre.
}
\label{eta_schw}
\end{figure}

\begin{figure}
\begin{center}
\includegraphics[width=\columnwidth]{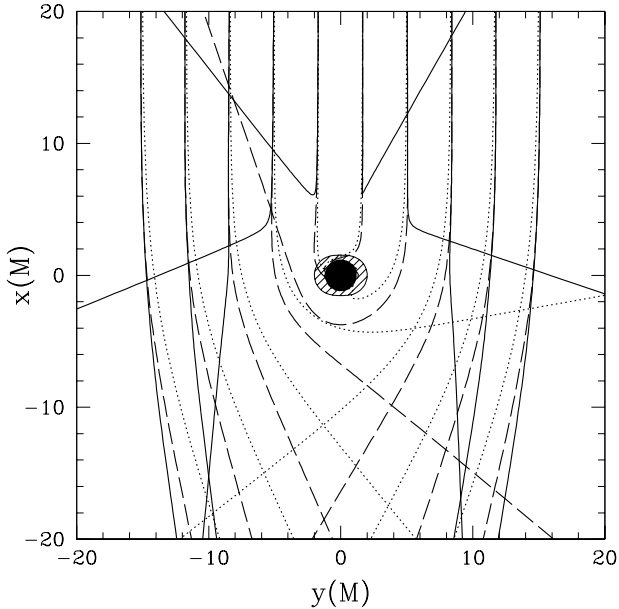}
\end{center}
\caption{ The paths of the ordinary and extraordinary polarisation eigenmodes
in the vicinity of maximally rotating Kerr black hole are shown by the dashed
and solid lines, respectively, for a number of impact parameters.  The dotted
lines show null geodesics for comparison.  The plasma parameters are the
same as those for figure \ref{eta_schw}.  In addition to the horizon, the
ergosphere is shown by the partially shaded region.  The rays originate from
$60^\circ$ above the equatorial plane.  The $y$ axis is orthogonal to the
the rotation axis of the black hole.
}
\label{eta_kerr}
\end{figure}

With general dispersion relation for cold magnetoactive plasmas,
equation (\ref{disp_gen}), and the ray equations, equations (\ref{ray_eqs}),
it is straightforward to explicitly construct rays.  The plasma geometry
outlined in the previous section will be used here as well, with the scales
set by $\omega_P(r=3M) = \omega_{\mbox{\tiny obs}}$ and
$\omega_B (r=3 M) = 2 \omega_{\mbox{\tiny obs}}$, where
$\omega_{\mbox{\tiny obs}}$ is the frequency observed at infinity.
In figure \ref{eta_schw} rays are propagated in the vicinity of a
Schwarzschild black hole.  For comparison, in figure \ref{eta_kerr} rays
are propagated near a maximally rotating Kerr black hole.  The null
geodesics are shown by the dotted lines for reference.  In both figures
the extraordinary mode (solid lines) is refracted the most, and the
ordinary mode (dashed lines) is refracted more than the null geodesics.
This is precisely what is expected on the basis of the capture cross sections
presented in \S\ref{PCCS}.  In addition to dispersive plasma effects,
comparison with the null geodesics demonstrates that general relativistic
effects are also significant.

\subsection{Intensity and Polarisation Maps} \label{IaPM}
\begin{figure}
\begin{center}
\begin{tabular}{cc}
\includegraphics[height=0.4\columnwidth]{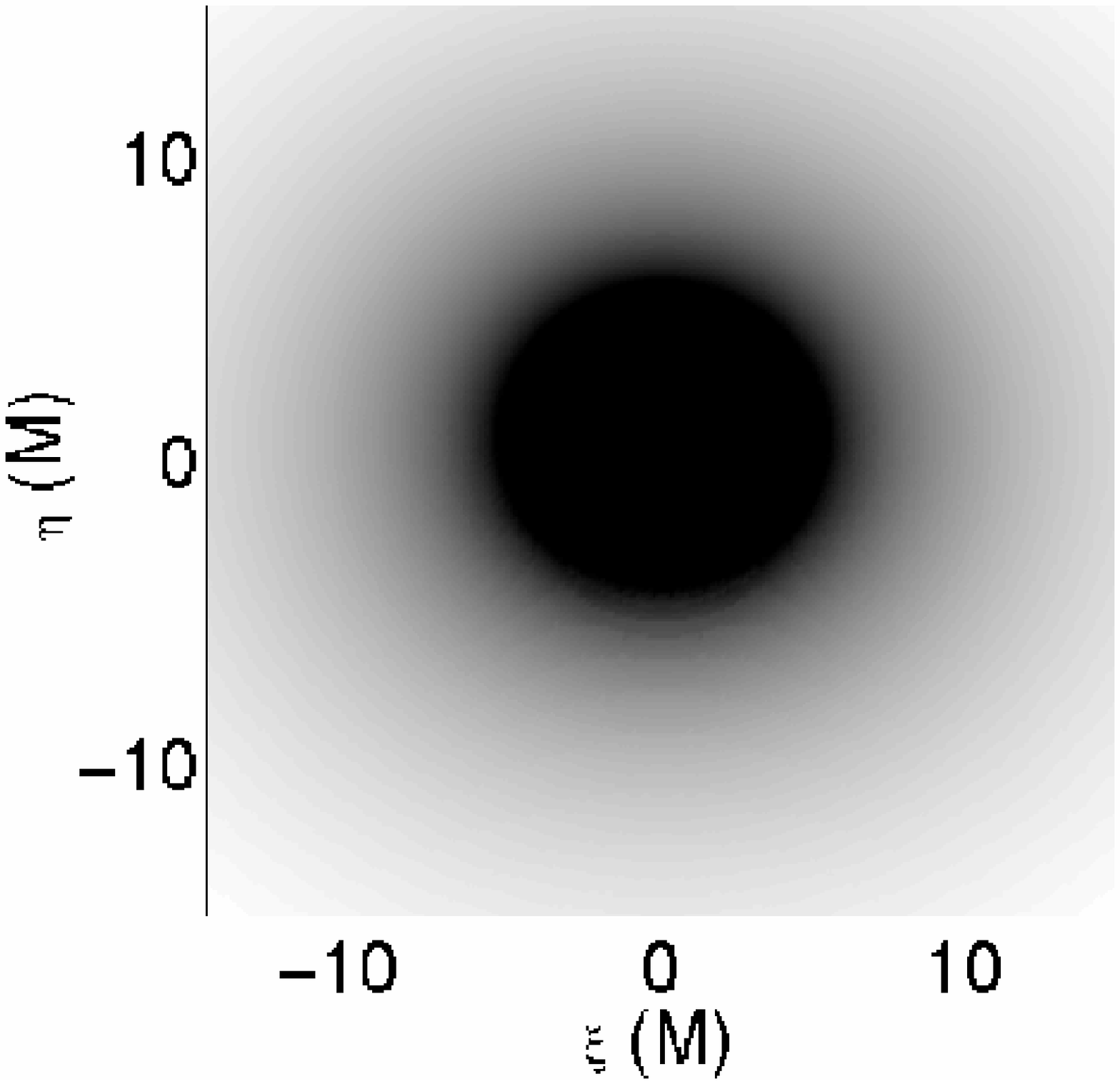} &
\includegraphics[height=0.4\columnwidth]{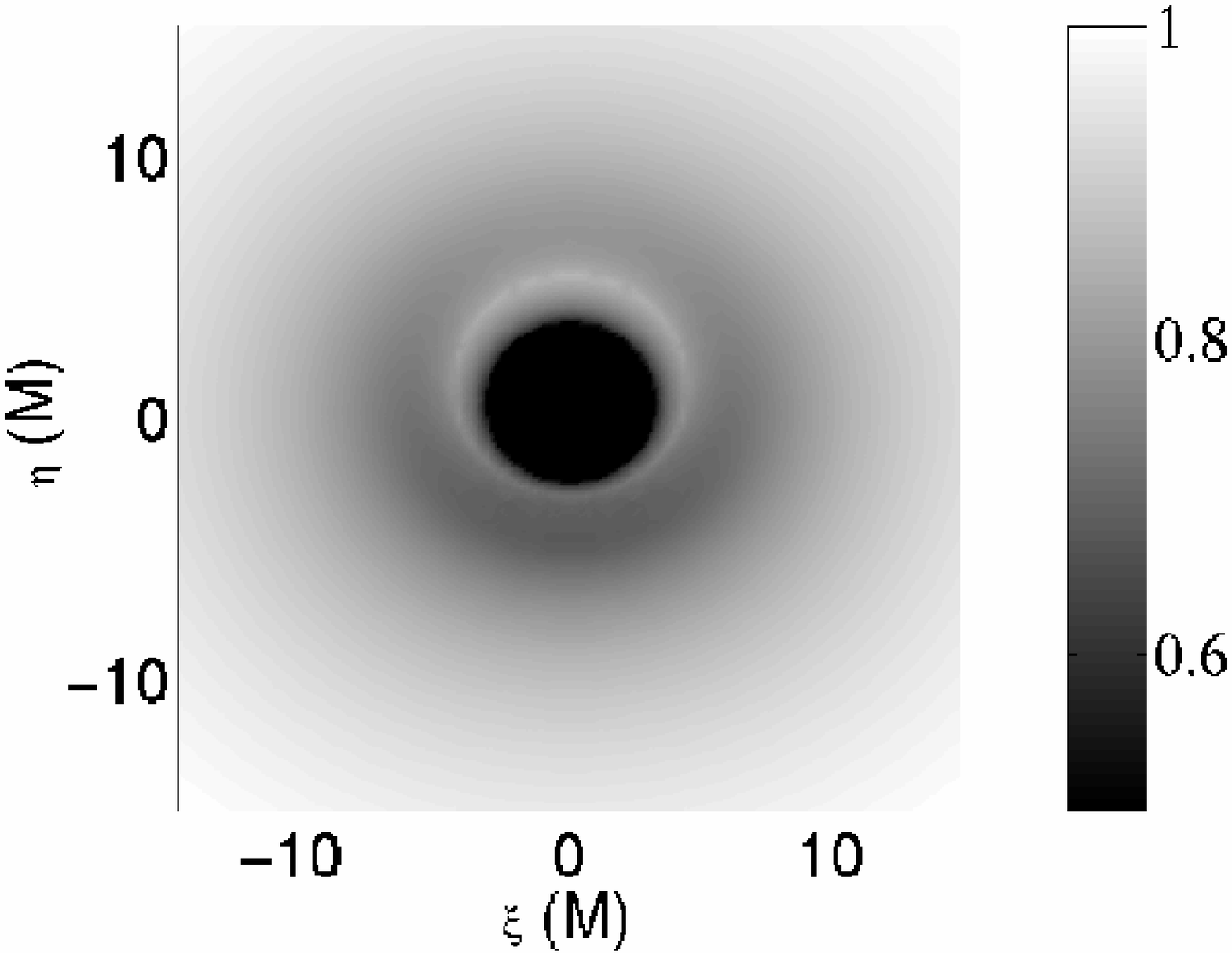} \\
(a) & (b) \\
\includegraphics[height=0.4\columnwidth]{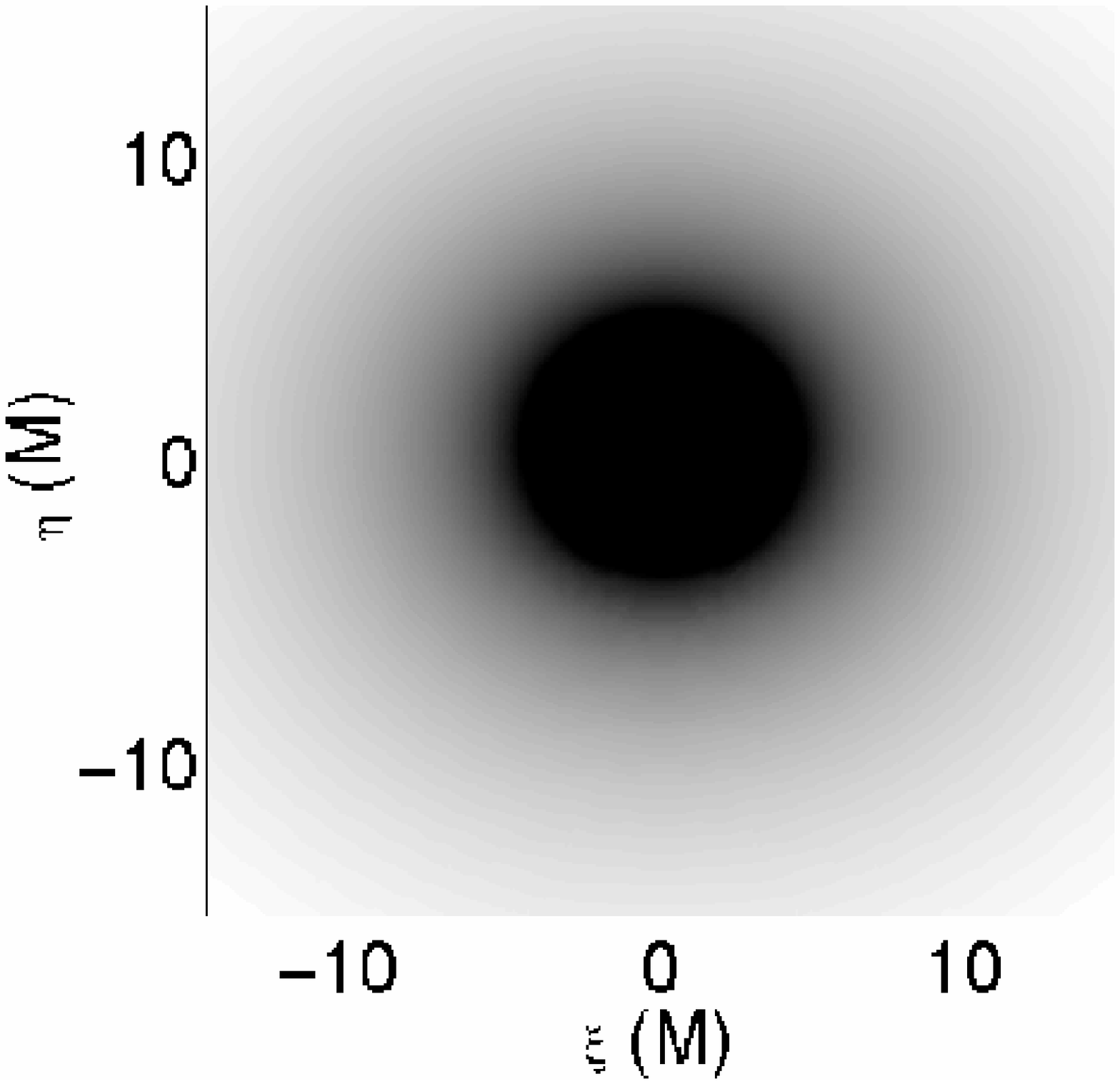} &
\includegraphics[height=0.4\columnwidth]{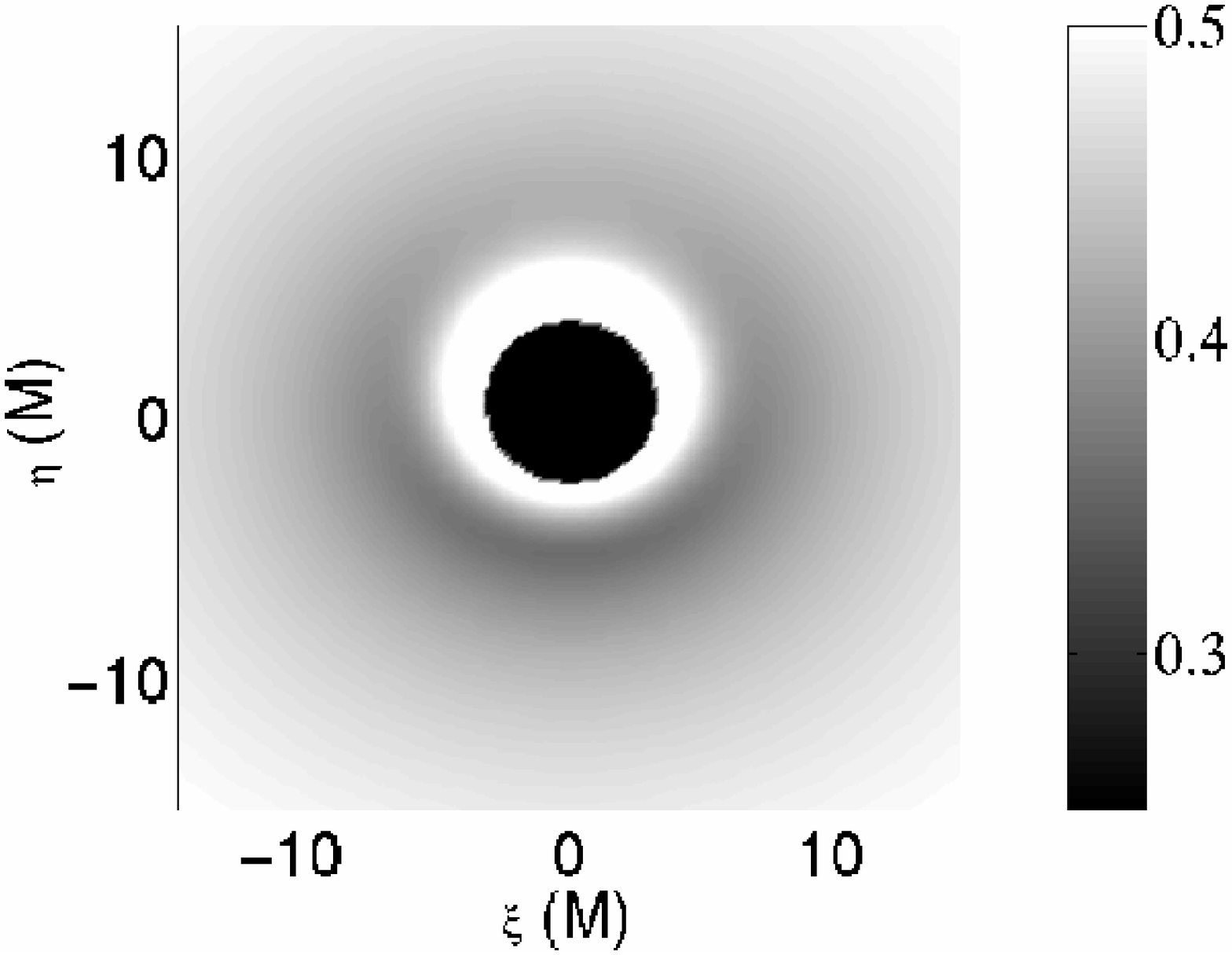} \\
(c) & (d) \\
\end{tabular}
\end{center}
\caption{Shown is the normalised intensity for an optically thick,
Shakura-Sunyaev disc around a Schwarzschild black hole
when (a) plasma effects are neglected, (b) plasma effects are included,
(c) only the left-handed circular polarisation (ordinary mode) is included,
(d) only the right-handed circular polarisation (extraordinary mode) is
included.  Note the different scales for total intensities ((a) and (b)) and
the polarised intensities ((c) and (d)).  The disk is inclined
$60^\circ$ relative to the line of sight.  $\xi$ is parallel to the equatorial
plane.  $\eta$ is in the line of sight-azimuthal axis plane.  The overall
scale is set by the choice of observation frequency and the parameters of
the disk and hence are not relevant here.  The plasma
geometry is the same as that for figures \ref{eta_schw} and \ref{eta_kerr}.
}
\label{maps}
\end{figure}
The impact that dispersive plasma effects can have upon the spectrum of an
accreting object can be illustrated by maps of the intensity.  Here, in 
addition to the plasma geometry employed in the previous two sections, an
optically thick Shakura-Sunyaev disk is introduced.  The emission
is solely from this disk and assumed to be thermal with
\begin{equation*}
T(r) \propto \left( 1-\sqrt{R_{\mbox{\tiny min}}/r} \right)^{3/10}
\end{equation*}
\citep[see \eg][]{Fran-King-Rain:92}.  The overall constant is dependent
upon a number of disk parameters and hence is not of particular interest
here.  Nonetheless, it is chosen such that
$kT(r=\infty) = \nu_{\mbox{\tiny obs}}$ for convenience.  The innermost
radius of the disk, $R_{\mbox{\tiny min}}$, is chosen to be $3M$.
Doppler effects due to the rotation of the disk are ignored here.

Shown in figure \ref{maps} are the intensity maps for when (a) plasma effects
are neglected, (b) plasma effects are included, (c) only the left-handed
circular polarisation (ordinary mode) is considered, (d) only the
right-handed circular polarisation (extraordinary mode) is considered.
Because the overall flux from the disk is dependent upon the details of
the accretion flow, the intensities are normalised by the highest intensity
in panel (b).  Comparing panels (a) and (b) demonstrates that including
dispersive plasma effects makes a significant difference.  This difference
originates primarily from contribution by the extraordinary mode shown in
panel (d).

As implied by figure \ref{cross_sections}, the shadow the black hole casts
upon the extraordinary mode is less than that cast upon the ordinary mode,
which is in turn less than that upon the null geodesics.  In addition to
the differences in the overall intensities, there is a substantial difference
between the contributions from the two polarisations as seen by comparing
panels (c) and (d).

\section{Conclusions} \label{Conclusion}
The covariant magnetoionic theory developed here is distinct in many respects
from the non-relativistic theory.  Firstly, it qualitatively changes the topology
of the dispersion relations, adding an entirely new branch, as shown in
\S\ref{BPF}.  Secondly, it allows the inclusion of gravitational lensing
effects, vital for application to compact accreting objects.  In addition,
as shown in \S\ref{PCCS} and \S\ref{TR}, dispersion due to plasma effects
can have a significant impact upon the propagation of photons in a dense
plasma environment near a black hole.  As demonstrated in \S\ref{IaPM},
this will lead to a modification of the spectrum.  As a result, studies
which neglect dispersive plasma effects may be inappropriate when the
observation frequencies are near the plasma and/or cyclotron frequencies.

On the other hand, because plasma effects have the capability of altering
the spectrum, it is possible for the underlying plasma to be observationally
probed using polarised flux measurements.  For example, if the horizon of
the black hole in the Galactic centre can be imaged
\citep[\cf][]{Falc-Meli-Agol:00}, observations of the
polarisation map could easily distinguish the nondispersive from the
dispersive case, placing limits upon the local magnetic field strength and
plasma density.  Integrated values for the polarisation could yield useful
information about the environments of other accreting systems, such as X-ray
binaries and pulsars.  Because these effects can be expected to be confined
to the decade in frequency surrounding the plasma frequency, they should be
easily distinguishable from the effects of different accretion models.

\section*{Acknowledgements}
The authors would like thank Eric Agol and Yasser Rathore for a number of
useful conversations and comments regarding this work.  This has been
supported by NASA grants 5-2837 and 5-12032.

\appendix
\section{Geodesic Motion in the Dispersion Formalism}
\label{massive_particles}
Given the dispersion relation in equation (\ref{deBroglie}),
\begin{equation*}
D(k_\mu,x^\mu) = k^\mu k_\mu - m^2 \,,
\end{equation*}
and the ray equations (\ref{ray_eqs}),
\begin{equation*}
\frac{dx^\mu}{d\tau} = \left( \frac{\partial D}{\partial k_\mu} \right)_{x^\mu}
\;\;\mbox{and}\;\;\;
\frac{dk_\mu}{d\tau} = -\left( \frac{\partial D}{\partial x^\mu} 
\right)_{k_\mu} \,,
\end{equation*}
it is possible to derive the geodesic equation.
The partial derivatives on the right side of the ray equations are
\begin{equation}
\left( \frac{\partial D}{\partial k_\mu} \right)_{x^\mu} = 2 k^\mu \,,
\end{equation}
and
\begin{align}
\left( \frac{\partial D}{\partial x^\mu} \right)_{k_\mu} &=
\left( \frac{\partial k_\alpha k_\beta g^{\alpha\beta}}{\partial x^\mu}
\right)_{k_\mu}
= k_\alpha k_\beta \frac{\partial g^{\alpha\beta}}{\partial x^\mu}
\nonumber \\
&= - k^\alpha k^\beta g_{\alpha\beta,\mu} \,.
\end{align}
Combining the ray equations gives
\begin{align}
\frac{d^2 x^\mu}{d \tau^2} &=
2 \frac{d k^\mu}{d \tau} = 2 \frac{d k_\nu g^{\mu\nu}}{d \tau}
\nonumber \\
&= 2 k_\nu \frac{d x^\alpha}{d \tau} 
\frac{\partial g^{\mu\nu}}{\partial x^\alpha}
+ 2 g^{\mu\nu} \frac{d k_\mu}{d \tau}
\nonumber \\
&= - 4 k^\beta k^\alpha g^{\mu\nu} g_{\beta\nu,\alpha}
+ 2 g^{\mu\nu} k^\alpha k^\beta g_{\alpha\beta,\mu}
\nonumber \\
&= - 4 k^\alpha k^\beta \frac12 g^{\mu\nu}
\left( g_{\alpha\nu,\beta} + g_{\beta\nu,\alpha} - g_{\alpha\beta,\nu} \right)
\nonumber \\
&= - \frac{d x^\alpha}{d \tau} \frac{d x^\beta}{d \tau} 
\Gamma^\mu_{\alpha\beta} \,,
\end{align}
where the definition of the Christoffel symbol was used,
$\Gamma^\mu_{\alpha\beta} \equiv \frac12 g^{\mu\nu} \left( g_{\alpha\nu,\beta}
+ g_{\beta\nu,\alpha} - g_{\alpha\beta,\nu} \right)$.  Collecting terms on
the left produces the well known geodesic equation:
\begin{equation*}
\frac{d^2 x^\mu}{d \tau^2} + \frac{d x^\alpha}{d \tau}
\frac{d x^\beta}{d \tau} \Gamma^\mu_{\alpha\beta}
= 0 \,,
\end{equation*}
or
\begin{equation*}
v^\nu \nabla_\nu v^\mu = 0
\;\;\mbox{where}\;\;
v^\mu \equiv \frac{d x^\mu}{d \tau} \,.
\end{equation*}

\bibliographystyle{mn2e.bst} \bibliography{cmt1.bib}

\bsp

\end{document}